\documentclass[fleqn,10pt]{wlscirep}
\usepackage[utf8]{inputenc}
\usepackage[T1]{fontenc}
\usepackage{bm}
\usepackage{comment}
\usepackage{upgreek}
\usepackage{soul}
\usepackage{enumitem}
\usepackage{graphicx}
\usepackage[font=footnotesize,labelfont=bf]{caption}
\usepackage{chngcntr}
\usepackage{hyperref}

\title{Learning Nonlinear Heterogeneity in Physical Kolmogorov-Arnold Networks}

\usepackage{authblk}

\author[1,$\ddagger$, $\dagger$]{Fabiana Taglietti}
\author[2,1,$\dagger$]{Andrea Pulici}
\author[3]{Maxwell Roxburgh}
\author[2]{Gabriele Seguini}
\author[4]{Ian Vidamour}
\author[5]{Stephan Menzel}
\author[1]{Edoardo Franco}
\author[6]{Michele Laus}
\author[4]{Eleni Vasilaki}
\author[2,*]{Michele Perego}
\author[7,*]{Thomas J. Hayward}
\author[1,8,*]{Marco Fanciulli}
\author[3,9,*]{Jack C. Gartside}

\affil[1]{Department of Materials Science, University of Milano-Bicocca, 20125 Milan, Italy.}
\affil[2]{National Research Council (CNR) - Institute for Microelectronics and Microsystems (IMM), Unit of Agrate Brianza, Via C. Olivetti 2, 20864 Agrate Brianza, Italy}
\affil[3]{Blackett Laboratory, Imperial College London, London, UK}
\affil[4]{School of Computer Science, University of Sheffield, Sheffield, UK}
\affil[5]{Forschungszentrum J\"ulich GmbH, PGI-7 Elektronische Materialien, J\"ulich, Germany.}
\affil[6]{Department of Science and Technological Innovation (DISIT), Universit\`a del Piemonte Orientale, 15121 Alessandria, Italy}
\affil[7]{School of Chemical, Materials and Biological Engineering, University of Sheffield, Sheffield, UK}
\affil[8]{Department of Chemistry, University of Turin, 10125 Turin, Italy}
\affil[9]{London Centre for Nanotechnology, Imperial College London, London, UK}

\affil[$\dagger$]{These co-first authors contributed equally}
\affil[*]{Corresponding author e-mails: j.carter-gartside13@imperial.ac.uk, marco.fanciulli@unito.it, t.hayward@sheffield.ac.uk, michele.perego@cnr.it}
\affil[$\ddagger$]{Present address: Forschungszentrum J\"ulich GmbH, PGI-7 Elektronische Materialien, J\"ulich, Germany.}

\begin{abstract}

Physical neural networks typically train linear synaptic weights while treating device nonlinearities as fixed. We show the opposite - by training the synaptic nonlinearity itself, as in Kolmogorov-Arnold Network (KAN) architectures, we yield markedly higher task performance per physical resource and improved performance-parameter scaling than conventional linear weight-based networks, demonstrating ability of KAN topologies to exploit reconfigurable nonlinear physical dynamics. 

We experimentally realise physical KANs in silicon-on-insulator devices we term `Synaptic Nonlinear Elements' (SYNEs), operating at room temperature, microampere currents, 2 MHz speeds and $\sim$750 fJ per nonlinear operation, with no observed degradation over $10^{13}$ measurements and months-long timescales.

We demonstrate nonlinear function regression, classification, and prediction of Li-Ion battery dynamics from noisy real-world multi-sensor data. Physical KANs outperform equivalently-parameterised software multilayer perceptron networks across all tasks, with up to two orders of magnitude fewer parameters, and two orders of magnitude fewer devices than linear weight based physical networks. These results establish learned physical nonlinearity as a hardware-native computational primitive for compact and efficient learning systems, and SYNE devices as effective substrates for heterogeneous nonlinear computing.

\end{abstract}

\begin{document}

\flushbottom
\maketitle
\thispagestyle{empty}

\section*{Introduction} 

\begin{figure}[t!]
\centering
\includegraphics[width=0.98\textwidth]{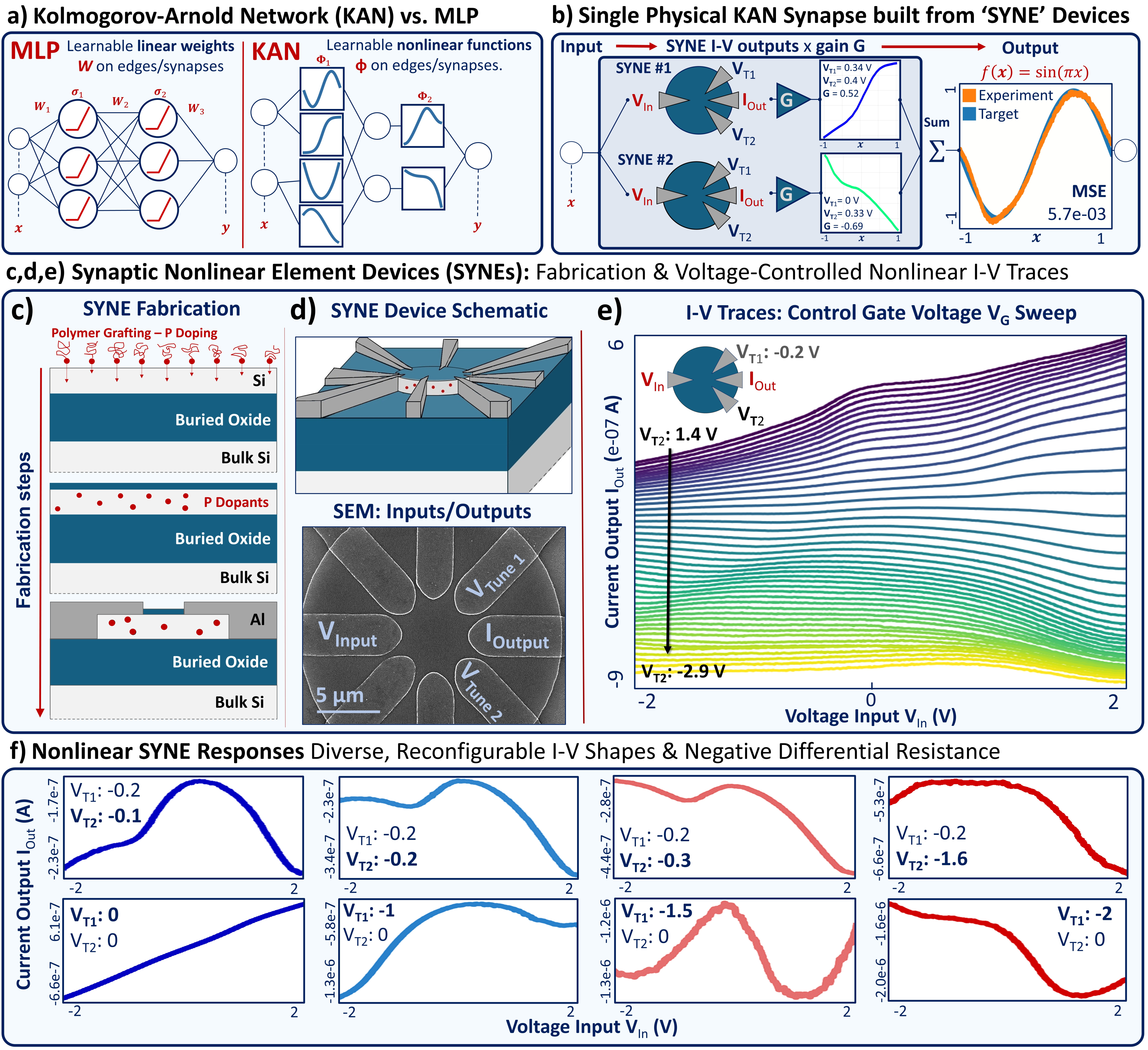}
\caption{\textbf{Kolmogorov-Arnold Networks and `Synaptic Nonlinear Element' (SYNE) Devices.} \textbf{a)} Comparison of multilayer perceptron (MLP) and Kolmogorov-Arnold network (KAN). MLPs employ relatively simple, identical nonlinear functions on all neurons (often ReLU), and learned linear weights on synapses/edges. KANs employ linear neurons that simply sum inputs, and complex synapses with learnable nonlinear functions.  \textbf{b)} Schematic of how a single KAN synapse is constructed by combining multiple physical SYNE devices in parallel. The input signal from a neuron is sent to $n$ SYNEs (here 2), each with learnable parameters which define the nonlinear function: two $V_\mathrm{Tune}$ control voltages which define the shape of the nonlinear $I$--$V$ response, input range scaling, and output gain $G$. The desired physical KAN synapse output is then given by a linear sum of the $n$ SYNE outputs, with a single bias term - here trained to produce a sine wave.  \textbf{c)} Fabrication process of SYNE devices. P-dopants are introduced to a 30 nm silicon-on-insulator layer via polymer-graft doping, then a 2\,nm SiO$_2$ oxide layer introduced via SC2 cleaning. Micron-scale disks are defined via maskless optical lithography then wet KOH etch. Al contacts are added via maskless optical lithography then thermal evaporation. \textbf{d)} SYNE device schematic and scanning electron micrograph. While eight contacts are patterned, only four are required for the scheme employed here: a voltage input $V_\mathrm{In}$, a current output $I_\mathrm{Out}$, and two tunable control voltages $V_\mathrm{Tune}$ (adjacent to the current output for maximal nonlinearity control). The remaining contacts are unused, providing scope for future interconnectivity and exploring additional control voltages. \textbf{e)} SYNE $I$--$V$ traces measured across a range of control voltages. $V_\mathrm{Tune 1}$ is held constant at -0.2\,V, $V_\mathrm{Tune 2}$ is swept 1.4\,V to -2.9\,V. \textbf{f)} SYNEs provide a broad, reconfigurable range of nonlinear $I$--$V$ responses including negative differential resistance - demonstrating the high expressivity afforded by a single SYNE device. Each row sweeps only a single control voltage, with the bottom row leaving the second control voltage fixed at 0\,V to highlight the range of nonlinear dynamics accessible with just a single control.}
\label{Fig1} 
\end{figure}

Physical learning systems offer rich, nonlinear dynamics and the potential for highly efficient processing\cite{markovic2020physics,momeni2025training,mcmahon2023physics,kudithipudi2025neuromorphic,sangwan2020neuromorphic,grollier2020neuromorphic,farmakidis2024integrated,kurebayashi2025technical,aguirre2024hardware,brunner2025roadmap}, but are typically constrained to small networks relative to software\cite{kudithipudi2025neuromorphic,momeni2025training,aguirre2024hardware,farmakidis2024integrated,muir2025road,brunner2025roadmap}. Despite impressive demonstrations, scaling remains a central challenge: fabrication complexity, device variability, and experimental noise impose harsh penalties as network sizes and device counts grow. Overcoming this is crucial to unlocking more powerful networks which present a real alternative to CMOS. 

One direction receiving less attention is developing new network architectures which directly exploit and learn the heterogeneous and programmable nonlinear dynamics in physical systems. To date, many physical neural networks have implemented software architectures optimised for GPUs/TPUs\cite{momeni2025training,markovic2020physics,aguirre2024hardware,farmakidis2024integrated,brunner2025roadmap}, which excel at linear matrix multiplications\cite{sutton2019bitter,hooker2021hardware,laydevant2024hardware}. These models are dominated by large arrays of learned linear weights on edges/synapses, paired with simple, fixed nonlinearity on neurons. A widespread example is the multilayer perceptron (MLP, Fig. \ref{Fig1}a), with most modern ML architectures including transformers\cite{lin2022survey,khan2022transformers} and convolutional neural networks (CNNs)\cite{gu2018recent} dominated by trainable linear weight matrices and static nonlinear neurons. This approach gains most of its computational power from sheer scale\cite{sutton2019bitter,hooker2021hardware,hestness2017deep,kaplan2020scaling, hua2025integrated,ahmed2025universal}, requiring vast and growing linear weight arrays. This leads to issues when transferring these architectures to physical networks, where increasing device counts hits hard experimental constraints on fabrication, device variability, and noise - severely limiting network size and performance\cite{kudithipudi2025neuromorphic,muir2025road,aguirre2024hardware,markovic2020physics,brunner2025roadmap}.

Such linear weight based networks also neglect some of the key strengths of physical systems: complex, heterogeneous nonlinearities which can be programmed and controlled on a per-device basis. While efforts continue to improve fabrication and develop new materials and devices with higher signal-to-noise, a complementary approach is developing network frameworks which are intrinsically `physics aware', and directly harness reconfigurable physical nonlinearity as a valuable computational resource. If these dynamics can be learned and leveraged, higher network performance per device count may be realised, or substantially reduced network sizes and device counts required to implement given tasks. Rather than engineering physical systems to implement linear weight array architectures developed for compatibility with linear, digital GPUs/TPUs, physics-based computing stands to benefit from architectural approaches with intrinsic synergy to devices exhibiting expressive, tunable nonlinear dynamics.

Computing schemes have been explored which make use of the diverse range of physical nonlinear dynamics, with encouraging results that highlight their potential as a computational resource. Physical reservoir and extreme-learning machine style schemes\cite{lee2024task,gartside2022reconfigurable,love2023spatial,allwood2023perspective,kurebayashi2025technical,markovic2020physics,liang2024physical,usami2021materio} often exploit a diverse range of nonlinear dynamics, and have demonstrated learning in overparameterised regimes at relatively small network sizes\cite{stenning2024neuromorphic}, and strong performance when training data is scarce\cite{ng2024retinomorphic}. However, these nonlinear dynamics are typically treated as fixed - initialised at the material/device fabrication stage, or varied as hyperparameters\cite{lee2024task,zolfagharinejad2025analogue,chen2020classification} and not learned (e.g. via gradient-based methods) to match the demands of a specific task. By not directly learning the locally-programmable nonlinear dynamics on offer in myriad physical systems, it is likely that powerful computational resources are being left unused. 

Recently, the `Kolmogorov Arnold Network'\cite{liu2024kan,arnol1957functions} (KAN, Fig. \ref{Fig1}a) architecture was proposed, which shifts learning from linear synaptic weights to trainable nonlinear synaptic functions. By refocusing learning onto per-synapse complex nonlinear functions, KANs have shown the ability to significantly reduce network size relative to MLP-style baselines\cite{liu2024kan,xu2024fourierkan,ss2024chebyshev,vaca2024kolmogorov,yang2024kolmogorov,bodner2024convolutional,somvanshi2025survey} across regression, vision, and time-series tasks including convolutional\cite{bodner2024convolutional} and transformer\cite{yang2024kolmogorov} inspired topologies. In physical neural networks, network size and parameter count correspond to both the number of devices which must be fabricated and the number of I/O control lines. Fabricating large physical arrays remains a key challenge, making network architectures potentially capable of reducing network size highly attractive. KANs can also improve interpretability, with interpretable ML an increasingly important concern\cite{molnar2020interpretable,rudin2022interpretable}, as the learned synaptic functions in a well-trained KAN tend to decompose the underlying nonlinear interactions in a dataset, which may then be directly inspected via the synaptic function shapes\cite{liu2024kan,somvanshi2025survey,vaca2024kolmogorov,yang2024kolmogorov}. Physics-based KANs have been proposed in theoretical studies\cite{peng2024photonic,stroev2025programmable,li2025fully} with promising results, but experimental realisations are crucial to confirming the potential of the approach. While translating the highly flexible mathematical nonlinear basis functions used in KANs (including splines, Fourier series \& Chebyshev polynomials) into the nonlinear dynamics of experimental physical systems represents a considerable challenge, we posit that the KAN architecture provides a natural framework for exploiting tunable heterogeneous nonlinear dynamics in physical neural networks, while reducing barriers to network size and scaleability. 



Here, we experimentally realise physical KANs, and show that learning heterogeneous nonlinear dynamics can substantially reduce network size and device count in physical learning systems. We implement KANs in doped silicon-on-insulator on-chip devices we term `Synaptic Nonlinear Elements' (SYNEs) to reflect their role in providing tunable synaptic nonlinearity. SYNEs are similar in design to `Dopant Network Processing Units' and `Reconfigurable Nonlinear Processing units' (DNPUs/RNPUs) demonstrated by van der Wiel et al\cite{chen_classification_2020,ruiz2020deep,ruiz2021dopant,zolfagharinejad2025analogue}, with adaptations to the material platform (introduction of thin SOI active layer), fabrication, and doping scheme which enable 2 MHz operation at room temperature. SYNEs operate at 0.1-1 microampere currents under $\pm$2 V input, giving $\sim$750 fJ per nonlinear operation (passing one datapoint through one device). SYNE devices are stable, with no degradation observed over $10^{13}$ measurements and months-long timescales (SI Fig. \ref{SI-device-stability}). We employ a time-multiplexed hardware-in-the-loop implementation in which physical nonlinearities are implemented by sequential experimental measurements of a single programmable SYNE device, with all nonlinear dynamics and experimental noise originate from the physical substrate. Linear scaling and summing operations are performed digitally off-chip. We form physical KAN synapses by combining SYNE devices in parallel (Fig. \ref{Fig1}b), and show that SYNEs provide a broad range of voltage-programmable, non-monotonic nonlinear transfer functions including negative differential resistance, enabling on-demand tuning of synaptic nonlinearity. After posting our preprint, we learned of a separate manuscript\cite{escudero2026physical} and accompanying thesis\cite{zolfagharinejad2025information} on physical KANs in electronic RNPU devices.

To learn the specific nonlinear physical dynamics required to perform given tasks, we employ a differentiable surrogate model of the SYNE device response; a data-driven digital twin\cite{manneschi2024noiseawaretrainingneuromorphicdynamic,wright2022deep,chen2022forecasting,ruiz2020deep} MLP trained on experimentally measured $I$--$V$ data (SI Fig. \ref{SI-Digi-Twin-Test}). The digital twin enables in-silico training in a hardware-in-the-loop pipeline, with control parameters learned via gradient descent on the digital twin transferred to experimental SYNE devices. We demonstrate learning of diverse nonlinear functions $f(x)$ by physical KAN synapses, including nested compositions $f(g(x))$ (Fig. \ref{Fig2}). We show that while a single SYNE synapse underperforms experimentally relative to software MLP baselines on function regression, interconnecting just two programmable physical synapses grants an emergent gain in performance and expressivity that inverts this relationship, outperforming software MLPs despite experimental noise. To quantify synapse and network expressivity (how diverse is the range of accessible nonlinear dynamics), we introduce a metric `Epsilon Expressivity', and show that it correlates strongly with function-representation performance (Fig. \ref{Fig3}) - enabling assessment of device configurations and architectures before costly training or fabrication runs.



 
Across nonlinear function regression (Figs \ref{Fig2} and \ref{Fig4}), classification (Fig. \ref{Fig5}a,b), and prediction of Li-Ion battery dynamics from noisy multi-sensor data (Fig. \ref{Fig5}c), physical KANs outperform parameter-matched software multilayer perceptron networks, requiring up to two orders of magnitude fewer trainable parameters, and up to two orders of magnitude fewer devices than physical linear weight based networks. We observe improved performance-parameter scaling relative to linear weight networks. Together, these results demonstrate that programmable physical synaptic nonlinearities can be reliably learned, composed, and deployed with an experimentally measured network with reduced trainable parameter and device counts relative to linear weight based networks. Our findings indicate that focusing learning and computation on programmable nonlinear activations can provide distinct advantages in physical learning networks, despite finite device expressivity and measurement noise. This work motivates Kolmogorov-Arnold architectures as a practical framework for exploiting reconfigurable nonlinear dynamics across diverse programmable nonlinear physical systems.


\section*{Results and Discussion}

\subsection*{Synaptic Nonlinear Element Devices (SYNEs)}

At the core of this work are the silicon-on-insulator devices that provide our reconfigurable nonlinear dynamics, which we term synaptic nonlinear elements (SYNEs). Shown in schematics and scanning electron micrograph in Figures \ref{Fig1}c) and d), SYNEs are P-doped (via polymer-graft) disk-shaped multi-terminal devices with 20\,$\upmu$m diameter fabricated in a thin $\sim$30~nm silicon-on-insulator layer with aluminium contacts, and a 6\,$\upmu$m diameter active region between the contacts. P-dopants are introduced to the top silicon layer via precision polymer doping \cite{perego_control_2018} to a concentration $\sim10^{18}$ cm$^{-3}$, depicted in the fabrication schematic in Figure \ref{Fig1}c). Devices are defined via maskless optical lithography and KOH wet-etch (details in methods)\cite{pulici_electrical_2023}. Here, four contacts are used: a voltage input $V_{\mathrm{in}}$, a current output $I_{\mathrm{out}}$, and two control electrodes biased with tuning voltages $V_{\mathrm{Tune1}}$ and $V_{\mathrm{Tune2}}$ which define the nonlinear shape of the $I$--$V$ response measured at $I_\mathrm{Out}$. The $V_\mathrm{Tune}$ contacts are selected adjacent to the output contact $I_\mathrm{Out}$, which was found to maximise tuneability of the nonlinear $I$--$V$ response. While four contacts are used, eight were fabricated, with the rest left floating and unused in this study. They are included in the device design to allow for exploration of employing additional tuning voltages, and future novel device-network connection topologies. Figure \ref{Fig1}e) shows an example of the range of nonlinear $I$--$V$ response shapes accessible by a SYNE device, at room temperature and ambient conditions. For each $I$--$V$ trace, $V_\mathrm{In}$ is swept from -2\,V to 2\,V in 0.01\,V steps while both $V_\mathrm{Tune}$ controls are fixed at constant values. 44 traces are shown, with $V_\mathrm{Tune 1}$ fixed at -0.2\,V for all, and $V_\mathrm{Tune 2}$ varied from 1.4\,V to -2.9\,V as the traces change from blue to yellow.

A broad range of complex nonlinear $I$--$V$ response shapes are observed, including positive, negative, and non-monotonic local slopes, both positive and negative $I_{\mathrm{Out}}$ values, and multiple turning points. As $V_{\mathrm{Tune2}}$ is made more negative, the transfer characteristics can smoothly transition into negative differential resistance (NDR), defined by a negative differential conductance $\mathrm{d}I_{\mathrm{Out}}/\mathrm{d}V_{\mathrm{in}} < 0$ over a finite bias window\cite{esaki1958new,esaki1966new}. The nonlinear responses accessible by the SYNE devices extend beyond those shown in Fig.~\ref{Fig1}e), where $V_{\mathrm{Tune1}}$ is held constant. This is illustrated in Fig.~\ref{Fig1}f), which shows two rows of $I$--$V$ traces in which one tuning voltage is fixed (top row: $V_{\mathrm{Tune1}}=-0.2~\mathrm{V}$; bottom row: $V_{\mathrm{Tune1}}=0~\mathrm{V}$) while the other tuning voltage is swept, revealing a diverse set of nonlinear behaviors including NDR, even when only a single tuning control is used and the other is left grounded (bottom row). 

Access to NDR is significant for physical Kolmogorov--Arnold network synapses as it enables intrinsically non-monotonic transfer functions with local negative slope and tunable turning points. This broadens the manifold of learnable edge nonlinearities available per physical device, increasing functional expressivity and potentially reducing the number of devices required to reach a given function approximation accuracy.

The physical mechanism behind the nonlinear transport and the emergence of NDR in SYNE devices likely reflects an interplay of bias-dependent conduction pathways and internal resistance modulation, and electrostatics in the active P-doped Si region (n-type) and at the interfaces between the active Si layer and Al contacts, and between the active Si and the buried SiO$_2$ oxide\cite{pulici_arxiv_2026}. A comprehensive characterization and model of the underlying physical transport mechanisms is an ongoing research activity and will be reported in the future.

\begin{figure}[t!]
\centering
\includegraphics[width=0.91\textwidth]{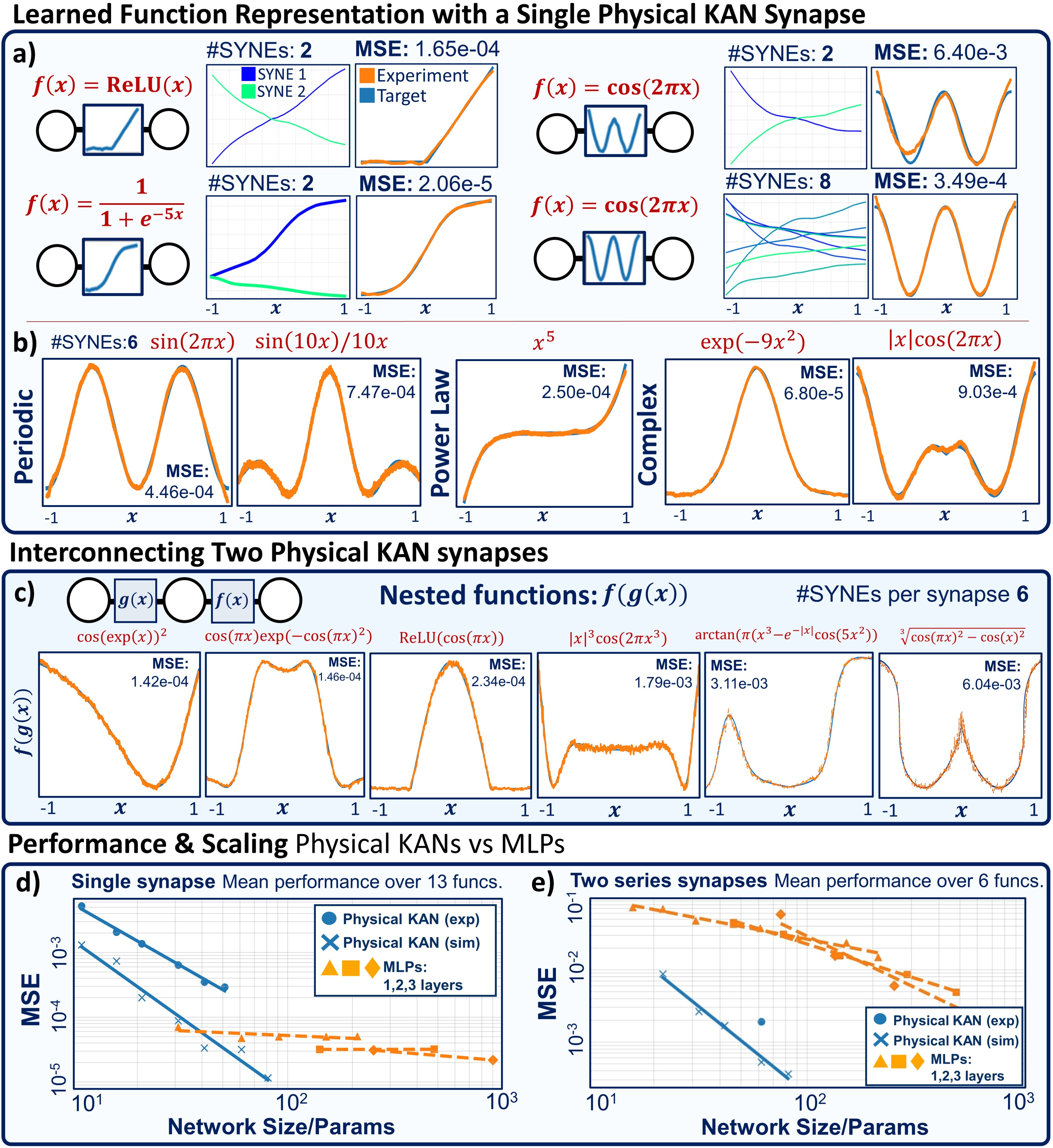}
\caption{\textbf{Learning Nonlinear Functions with Physical KAN Synapses} \textbf{a)} Nonlinear functions are learned in-silico and experimentally realised using a single physical KAN synapse. Using a differentiable digital twin model of a SYNE device, backpropagation is used to learn the experimental parameters required to experimentally reproduce arbitrary functions via SYNE output currents. The blue/cyan traces on the second and fifth column plots show the output currents of each SYNE in the synapse. The orange trace in the third and sixth columns is the experimentally-measured output of the KAN synapse, and the blue trace the ground truth target. In this work, a single SYNE device is employed - measured sequentially with different $V_\mathrm{Tune}$ values to implement a larger network, with output gain and linear summation performed digitally off-chip. Using only two SYNEs, simple ReLU and sigmoid functions are implemented at high accuracy. For functions with higher harmonic content such as cos$(2 \pi x)$, a greater number of SYNE devices per synapse can be employed to improve accuracy - demonstrated here by two SYNEs which implement a lower-accuracy representation, and eight SYNEs achieving a higher accuracy. \textbf{b)} A broad range of functions are experimentally accessible using a single KAN synapse containing 6 SYNE devices, measured at 2 MHz per datapoint with 10k datapoints per function. \textbf{c)} Two KAN synapses are connected in series, experimentally realising complex nested $f(g(x))$ functions. Series interconnection enables more expressive synapses. \textbf{d,e)} MSE vs. network size/trainable parameters averaged over multiple nonlinear functions, comparing experimental and simulated physical KAN synapses, and MLPs. For a single KAN synapse (d), performance is averaged over 13 functions, including the 8 functions in a) (details in SI). Here, software MLPs outperform experimental physical KANs, although simulated (digital twin) physical KANs beat MLPs. For two series KAN synapses (e), performance is averaged over the 6 functions shown in b). Here, experimental and simulated physical KANs beat MLPs due to the enhanced expressivity of series-connected synapses.}
\label{Fig2} 
\end{figure}

\subsection*{Physical Kolmogorov-Arnold Synapses from SYNE Devices}

While the range of accessible nonlinear functions from a single SYNE device is broad, it does not provide the arbitrary function representation for KAN synapses/edges that a B-spline or alternative basis (e.g. Fourier series\cite{xu2024fourierkan}, Chebyshev polynomials\cite{ss2024chebyshev}) does in software. To solve this, we construct physical KAN synapses from a number $n$ of SYNE devices in parallel, as shown in Figure \ref{Fig1}b), via a time-multiplexed scheme with off-chip linear operations as described above. Each SYNE device within a synapse has five trainable parameters: two $V_\mathrm{Tune}$ voltages, a linear gain $G$ (restricted to a max of $3\times$ for reasonable implementation by efficient on-chip amplifiers), and the minimum and maximum input voltages $V_\mathrm{In~Min}$, $V_\mathrm{In~Max}$. Neurons are linear, and perform summation of incoming information from synapses with a trainable linear bias term $c$. Incoming information from a neuron is sent to the $n$ SYNEs of a synapse, which provide different nonlinear shapes via different $V_\mathrm{Tune}$ values, with $n = 2$ in Figure \ref{Fig1}b). The output currents $I_\mathrm{out}$ of each SYNE are then relatively scaled by a gain $G$ (implemented digitally off-chip here, see methods for projected circuit descriptions using an amplifier per SYNE or an amplifier per hidden/output neuron with a VCM memristor\cite{dittmann2021nanoionic} per SYNE providing relative output scaling). The scaled outputs are then linearly summed (implemented digitally off-chip here, Kirchoff summing in projected circuits) to produce the KAN synapse output. Just two SYNE devices are sufficient to realise a wide range of nonlinear functions at high accuracy, as demonstrated in the $\sin(\pi x)$-function shown in Figure \ref{Fig1}b) with a mean square error (MSE) of 5.7e-3. The required $V_\mathrm{Tune}$ and gain $G$ values for implementing an arbitrary nonlinear response are learned via backpropagation of a differentiable digital twin model (an MLP trained on experimental $I$--$V$ data) and then transferred to the experimental device for measurement. For clarity, we reiterate that here a single SYNE device is employed in a time-multiplexed scheme to sequentially represent the $n$ SYNEs in a synapse. All voltages, currents, and nonlinear function shapes are applied and measured experimentally, with gain and linear summing applied digitally off-chip. As the physical devices are refined in future to provide maximal nonlinear expressivity, we foresee that the number of SYNE devices per synapse required to reach a specific level of expressivity or performance may be reduced.


\subsection*{Function Representation with Physical KAN Synapses}

Figure \ref{Fig2} demonstrates the ability of SYNE-based physical KAN synapses to learn and represent a broad range of arbitrary functions. We consider a range of univariate nonlinear functions $f(x)$ where $-1 \le x \le 1$. We construct a mathematically-defined target function of 1k points, and learn the required SYNE parameters to reproduce the function via backpropagation of the digital twin. We then experimentally transfer these parameters to physical SYNE devices and evaluate them over 10k points, and assess performance via MSE. Figure \ref{Fig2}a) shows the results for three $f(x)$ functions: ReLU$(x)$, $\frac{1}{1 + e^{-5x}}$ (sigmoid $5x$), and $\cos(2 \pi x)$. For each function, we show a schematic of a single SYNE synapse with the experimentally-realised function displayed between an input and output neuron. In the panels to the right headed \#SYNEs, the blue/cyan traces show the experimentally-measured output currents of each SYNE device, scaled by the gain $I_\mathrm{Out} * G$. The panels headed MSE: show the experimentally-measured synapse output (orange), linearly summing the separate SYNE responses, against the ground-truth target function (blue). For the ReLU and sigmoid $5x$ functions, two SYNEs provide a high-accuracy representation with MSEs of 1.65e-4 and 2.06e-5 respectively. For functions with higher frequency content and multiple turning points, such as $\cos(2 \pi x)$, two SYNEs per synapse provides a reasonable representation at an MSE of 6.4e-3, but fails to capture certain details of the target function. This can be addressed by increasing the number of SYNEs per synapse to provide greater expressivity, similar to increasing the number of terms in a Fourier series. The bottom right plot of Figure \ref{Fig2}a) shows a $\cos(2 \pi x)$ function implemented by a synapse containing 8 SYNEs, with an increased quality representation and lower MSE of 3.49e-4. The number of SYNEs per synapse may be varied to trade off between network size and energy, against expressivity and performance - and may be optimised at a per-synapse level dependent on the required function complexity on each synapse. Figure \ref{Fig2}b) shows learned experimental representations of five further $f(x)$ functions with a range of oscillatory, asymmetric and symmetric shapes using a single physical KAN synapse containing 6 SYNEs, highlighting the flexibility of these devices for function representation and their suitability for providing the reconfigurable nonlinearity required by Kolmogorov-Arnold synapses.

Beyond considering single physical KAN synapses, we can connect the output of one synapse to the input of another, providing a framework for representing more complex nested $f(g(x))$ function forms or arbitrary functions with high frequency content that cannot be represented using a single synapse. In Figure \ref{Fig2}c) we demonstrate this, experimentally implementing a range of six functions of the form $f(g(x))$, including more complex shapes and sharper curvature than the functions considered in figs. \ref{Fig2}a,b).

We now compare the relative function representation performance of SYNE-based KAN synapses in simulation (via digital-twin models) and experiment, against software MLP networks. Figure \ref{Fig2}d) shows the mean MSE performance against the network size/trainable parameter count over a range of 13 nonlinear functions (including the 8 shown in figs. \ref{Fig2}a,b), details and function list in SI) for a single KAN synapse (2-12 SYNEs per synapse) and MLPs (1-3 hidden layers, 50-300 neurons per layer). Here, even single layer MLPs outperform an equivalently parameterised experimental physical KAN synapse, although a simulated physical KAN synapse which is not subject to experimental noise or model-reality gap during transfer outperforms MLPs at higher SYNE-counts. While the scaling exponents of the parameter-performance curves of the experimental physical KAN synapses are substantially steeper than those of the MLPs - simply growing the KAN network size to improve relative performance runs counter to the vision of compact, highly-expressive physical neural networks which maximally exploit reconfigurable physical nonlinearities. 


Instead, we can improve performance via architecture. By rearranging SYNE devices into series-interconnected synapses, we show that experimental performance beyond software MLPs is attainable. Figure \ref{Fig2}e) shows the same MSE vs. network size plot for two KAN synapses in series, averaged over the six functions shown in Fig. \ref{Fig2}c). Here, both simulated and experimental physical KANs outperforming software MLPs at equivalent parameter counts (experimental data is for 6 SYNEs per synapse only, separate plots for each $f(g(x))$ function in SI Fig. \ref{SI-f-g-x-function-regression}). 

This result is significant for physical KANs and other schemes focusing computation on programmable nonlinearity. It is challenging to engineer an experimental device that can match the arbitrarily-high expressivity and noise-free precision of a mathematically-defined spline in software, especially within the constraints of making systems compact, chip-compatible and efficient. Here, we show simply connecting two devices in series enhances expressivity sufficiently to outperform software MLPs. Moving to deeper/series synapses is perhaps a clear step forwards, but the boost in performance is an important result - compounding experimental noise, device imperfections and enhanced complexity of training when going deeper all work against system performance. This result shows that physical systems of lesser expressivity relative to software splines and other KAN basis functions are able to realise simple modular architectures where interconnected devices can be considered a single `meta synapse', enhancing performance via optimising device-architecture synergy.


\begin{figure}[t!]
\centering
\includegraphics[width=0.94\textwidth]{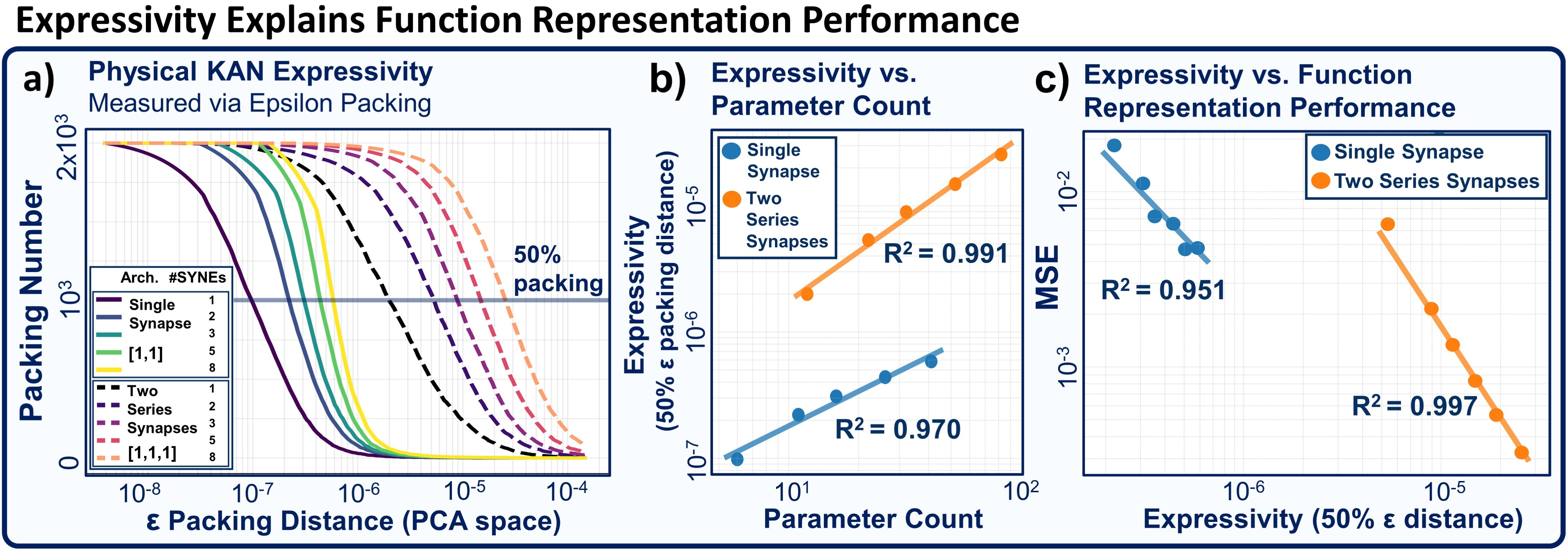}
\caption{\textbf{Expressivity via Epsilon Packing is Strongly Correlated with Function Representation Performance.} \textbf{a)} Physical KAN synapse expressivity is quantified for single and two series-connected synapses comprising 1-8 SYNEs, via Epsilon Packing. 2000 curves are generated for each synapse by sweeping all SYNE parameters, and embedded in a PCA space. The packing number denotes how many curves are at least a distance $\epsilon$ apart from their nearest neighbour in PCA space, e.g. how varied/expressive is the set of possible curves. The packing number at a given $\epsilon$ increases as more SYNEs are added to a synapse. Moving to two series synapses substantially increases expressivity, explaining why physical KANs move from underperforming relative to MLPs for a single synapse, to beating them for two series synapses. The faint horizontal line denotes the $\epsilon$ distance at which 50\% of total curves can be packed. \textbf{b)} To allow us to quantitatively assess the link between expressivity metricised via epsilon-packing, we extract a single value from each of the traces in (a) - the packing distance $\epsilon$ at which 50\% of curves can be packed. We plot this here against parameter count, showing a strong positive correlation (power law fit) with a steeper gradient and higher absolute expressivity for the two series synapse case. \textbf{c)} We now explore correlation between the expressivity (via the 50\% packing $\epsilon$ distance) and the mean learnt function representation MSE performance across a set of ten complex nonlinear functions (methods). The expressivity is strongly correlated with the MSE performance, with a steeper gradient in the two series synapse case. This result highlights the efficacy of using expressivity, quantified simply and computationally cheaply via epsilon packing in PCA space, as a metric for assessing the power of a given device or system for learning diverse ranges of complex nonlinear dynamics. 
}
\label{Fig3} 
\end{figure}

\subsection*{Synaptic Expressivity and Performance}

To quantify the expressivity of our physical KAN synapses, and compare the benefits of different synaptic architectures, we employ an approach termed `epsilon packing'\cite{kolmogorov1959varepsilon} to produce a quantitative expressivity metric, `Epsilon Expressivity', that directly measures the diversity of achievable transfer-curve shapes. We show that the quantified expressivity is highly correlated with the learned function regression performance across a diverse range of challenging functions (Fig. \ref{Fig3}). Figure \ref{Fig3}a) shows the results of epsilon expressivity assessed on a range of single [1,1] and two series-connected [1,1,1] synapses. Synapses are considered with 1-8 SYNEs each. 


\begin{figure}[t!]
\centering
\includegraphics[width=0.94\textwidth]{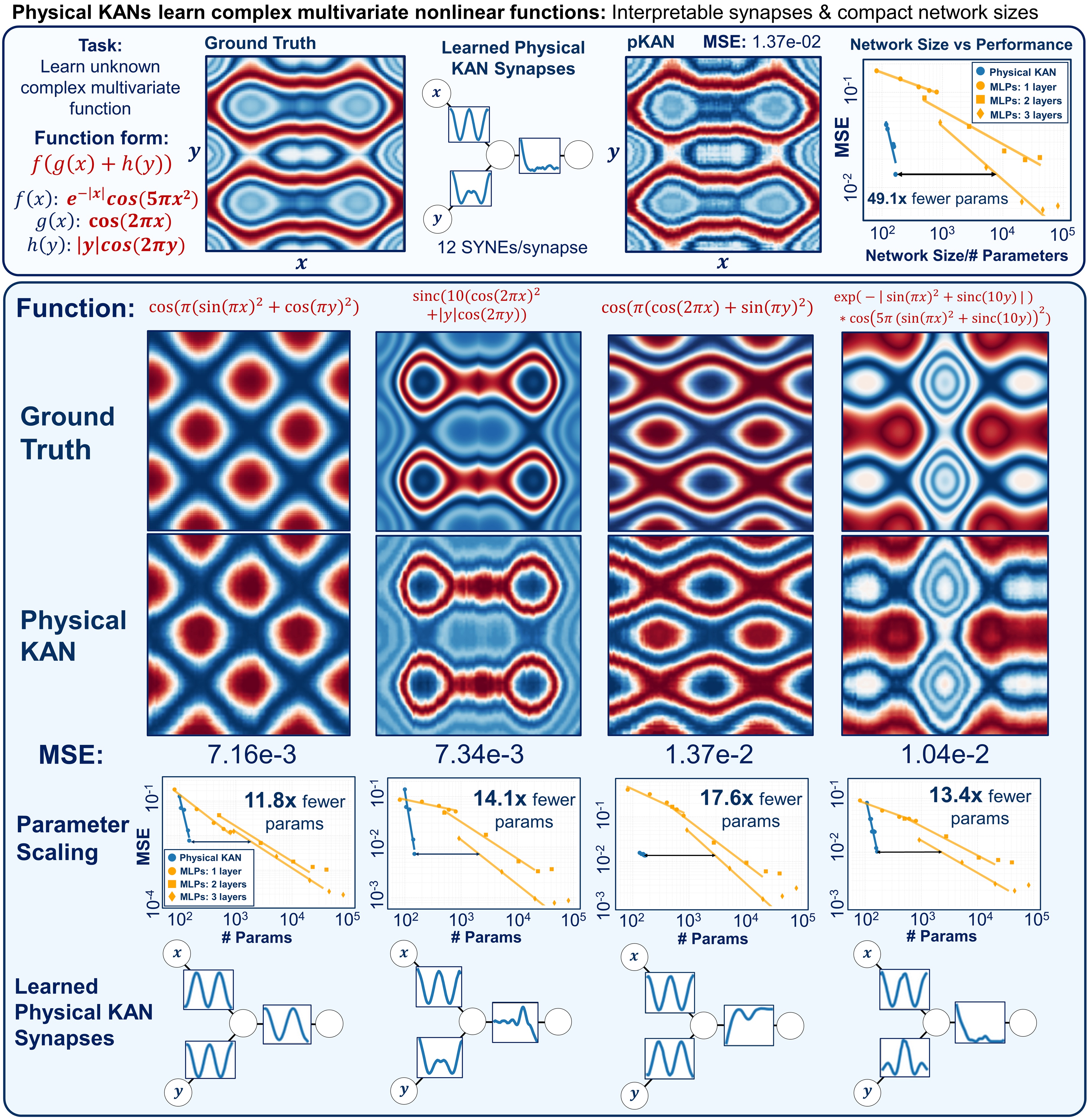}
\caption{\textbf{Learning Multivariate Nonlinear Functions with Physical KANs.} Better performance and smaller network sizes are found relative to software multilayer perceptrons, with interpretability from learned synaptic shapes. \textbf{a)} We implement a set of 2D $xy$ regression tasks on nonlinear functions with the form $f(g(x)+h(y))$ where $f(x), g(x), h(y)$ are unknown independent functions. A training set is composed of 80 $\times$ 80 points from -1 to 1 in $x$ and $y$, and a [2,1,1] physical KAN network with 12 SYNE devices per synapse is trained via backpropagation on the digital twin model to reproduce the function output for 800 epochs. The physical KAN network has no prior knowledge of the function composition, but on inspection of the learned synaptic shapes has learned to approximate the underlying function forms of the three separate $f,g,h$ functions - aiding symbolic interpretability. The learned parameters are then experimentally transferred to the hardware SYNE devices, and evaluated on a test set comprising 100 $\times$ 100 $x,y$ points. The `pKAN' (physical KAN) heatmap shows the experimental test set, with an MSE of 1.37e-02 and good reproduction of function features including high-frequency components at the edges and corners. Performance is compared between physical KANs (experimental) and software multilayer perceptron networks at a range of network sizes/trainable parameter counts. Physical KAN network sizes are reduced by selectively pruning SYNE devices,  achieving better performance than equivalently parameterised multilayer perceptrons which require 49.1$\times$ more parameters than physical KANs to achieve equal performance. \textbf{b)} Four more functions are learned and experimentally tested on physical KANs. In each case, physical KANs outperform MLPs at higher parameter counts, with approx. 30$\times$ reduction in network size relative to equivalently-performing multilayer perceptrons. Learned synaptic shapes aid interpretability of the underlying functions and work well in most cases, but are not without errors such as the imperfect cos($\pi x$) in the final layer synapse in the bottom left network plot.
}
\label{Fig4} 
\end{figure}

Calculating epsilon expressivity has two parts: nonlinear curve generation, then expressivity assessment. For each synapse/SYNE combination, 2000 $I$--$V$ curves are generated by sparsely sampling all available control parameters. These curves are then embedded as points within a 32-dimensional principal component analysis (PCA) space\cite{jolliffe2011principal} to form a point cloud. We then employ the concept of epsilon packing (which in a nice loop of history originates from foundational work by Kolmogorov\cite{kolmogorov1959varepsilon} amongst others\cite{kegl2002intrinsic}) to quantify the expressivity, or breadth and diversity of non-degenerate nonlinear curves within the PCA space point cloud. The concept is to take a fixed distance $\epsilon$ in PCA space ($\epsilon$ is the x-axis of Fig. \ref{Fig3}a)), and find how many of your nonlinear curves are at least a distance $\epsilon$ away from their nearest neighbour. The number of curves at least $\epsilon$ apart is termed the `packing number' (y-axis of Fig. \ref{Fig3}a)). The process is iterative, gradually increasing the size of $\epsilon$ until none of the considered curves are further than $\epsilon$ apart, or until a desired $\epsilon$-threshold is reached. 

Figure \ref{Fig3}a) shows the results of this process. For small $\epsilon$ values, all considered synaptic architectures pack all 2000 curves. As $\epsilon$ increases, less expressive synaptic architectures fail to pack all curves. Architectures consisting of a single synapse of parallel SYNEs fail to pack first, beginning with those lower SYNE counts, then proceeding to higher counts. The two series-connected KAN synapses cases are seen to retain full packing of all 2000 curves to much larger $\epsilon$ distances, demonstrating the increased expressivity and diversity of the representable nonlinear functions granted by moving to deep/interconnected synapses. To compress and extract the information in the epsilon packing plot, we define a threshold $\epsilon$ value where 50\% (1000) of the total curves meet the packing criteria, denoted by the light horizontal line. We term this extracted value the `epsilon expressivity' of the system. Figure \ref{Fig3}b) plots this epsilon expressivity against the total parameter count (5 parameters per SYNE + 1 per hidden/output neuron), with strong correlation (power law fits) and substantially larger absolute values and scaling exponents observed for the two series-connected synapse case, clearly displaying the benefits of spending a finite device/parameter count on interconnected/deep synaptic topologies. Finally, in Figure \ref{Fig3}c) we evaluate the explicit link between epsilon expressivity values and MSE performance on learned function regression tasks for single and series synapses (1-8 SYNEs per synapse), considering mean performance over a set of 10 nonlinear $f(x)$ and $f(g(x))$ functions (details in SI). We see strong correlation between expressivity and performance - with power law fits returning $R^2$ values of 0.997 and 0.951 for the two series-connected and single synapse cases respectively. 

This is a substantial result, as sampling the possible curve sets and calculating PCA-space epsilon packing is computationally lighter and orders of magnitude faster than assessing performance over multiple regression functions and trials - with the result shown here demonstrating that high expressivity scores can be taken as strong indicators of improved performance. This has high utility - for instance, to efficiently assess operational schemes and control/parameter choices, such as determining the optimal trade-off between number of control inputs/voltages shaping device nonlinearity or determining the optimal architecture in which to arrange a finite number of devices for best performance. As new candidates emerge as potential hardware primitives for physical networks driven by reconfigurable nonlinear dynamics, we propose epsilon expressivity as a tool for the community to directly assess and compare between the different physical systems being explored (including potentially emerging software basis functions in addition to physical devices). In the SI we discuss the scarcity of existing metrics for quantifying nonlinear expressivity, where existing approaches fall short, and show that once the nonlinear curve set is generated and embedded in PCA space a number of other quantitative metrics may be employed, including `soft diameter' of the point-cloud in PCA space.

\subsection*{Learning Multivariate Nonlinear Functions}

So far, we have considered learning univariate one-dimensional nonlinear functions. These are the foundational building block of KANs which use learnable nonlinear activations as their computational primitives. The SYNE synapses may now be interconnected into larger architectures to learn to solve tasks which a one-dimensional nonlinear basis (e.g. a single synapse, spline function, etc) cannot.

In Figure \ref{Fig4}, we evaluate function regression performance on multivariate nonlinear functions of the form $f(g(x)+h(y))$ which are nested and composite, for example the function $f(x) = e^{-|x|\cos(5 \pi x)}$, $g(x) = \cos(2 \pi x)$, $h(y) = |y|\cos(2 \pi y)$ shown in Figure \ref{Fig4}a). We employ a physical KAN architecture of [2,1,1] with 12 SYNEs per KAN synapse: two input neurons corresponding to $x$ and $y$ values, connected to a single hidden neuron, which is connected by a synapse to a final output neuron. This architecture provides the network with some inductive bias as it is constructed matching the form of our $f(g(x)+h(y))$ functions, but allows us to evaluate whether the learned nonlinear functions of physical KANs are able to learn and represent and decompose the underlying nonlinearities of the composite $f,g,h$ functions, from unknown data with no prior knowledge of the function forms. We start by constructing a training set by sampling $f(g(x)+h(y))$ on a grid of 80 $\times$ 80 points between -1 and 1 for $x$ and $y$ (see the `Ground Truth' panel in Fig. \ref{Fig4}a)), and training to minimise MSE loss via backpropagation over 800 epochs. We then examine the learned nonlinear synaptic functions (see the `Learned Physical KAN Synapses' panel). Examining these function shapes, $g(x) = cos(2\pi x)$ and $h(y) = |y|\cos(2 \pi y)$ are shown in Fig. \ref{Fig2}b), we see that the physical KAN has indeed learned to represent the forms of the individual $f,g,h$ functions, rather than finding an alternative form that also results in low MSE loss. We now evaluate function regression performance by experimentally inputting a 100$\times$100 grid of $x,y$ points between -1 and 1 and plot the measured values at the output neuron. The results of this test set experiment are shown in the `pKAN' heatmap panel, with an MSE of 1.37e-2 against the mathematical ground truth. The function here is complex, with a range of high- and low-frequency oscillations and nested positive and negative regions, which are reproduced by the physical KAN output. We compare MSE performance vs. network size for a range of 1-3 hidden layer MLPs and physical KANs where SYNE devices have been iteratively pruned from synapses (methods), and find that the best performing physical KAN requires 49.1$\times$ fewer trainable parameters than an equivalently performing MLP.

We repeat this process for four more multivariate nonlinear functions, shown in Figure \ref{Fig4}b). In each case, the physical KANs are able to learn the functions, providing interpretability via the learned nonlinear synaptic functions and outperforming equivalently parameterised MLPS, with $\sim$11.8-17.6$\times$ reduction in parameter count/network size relative to equally-performing MLPs.

\begin{figure}[t!]
\centering
\includegraphics[width=0.94\textwidth]{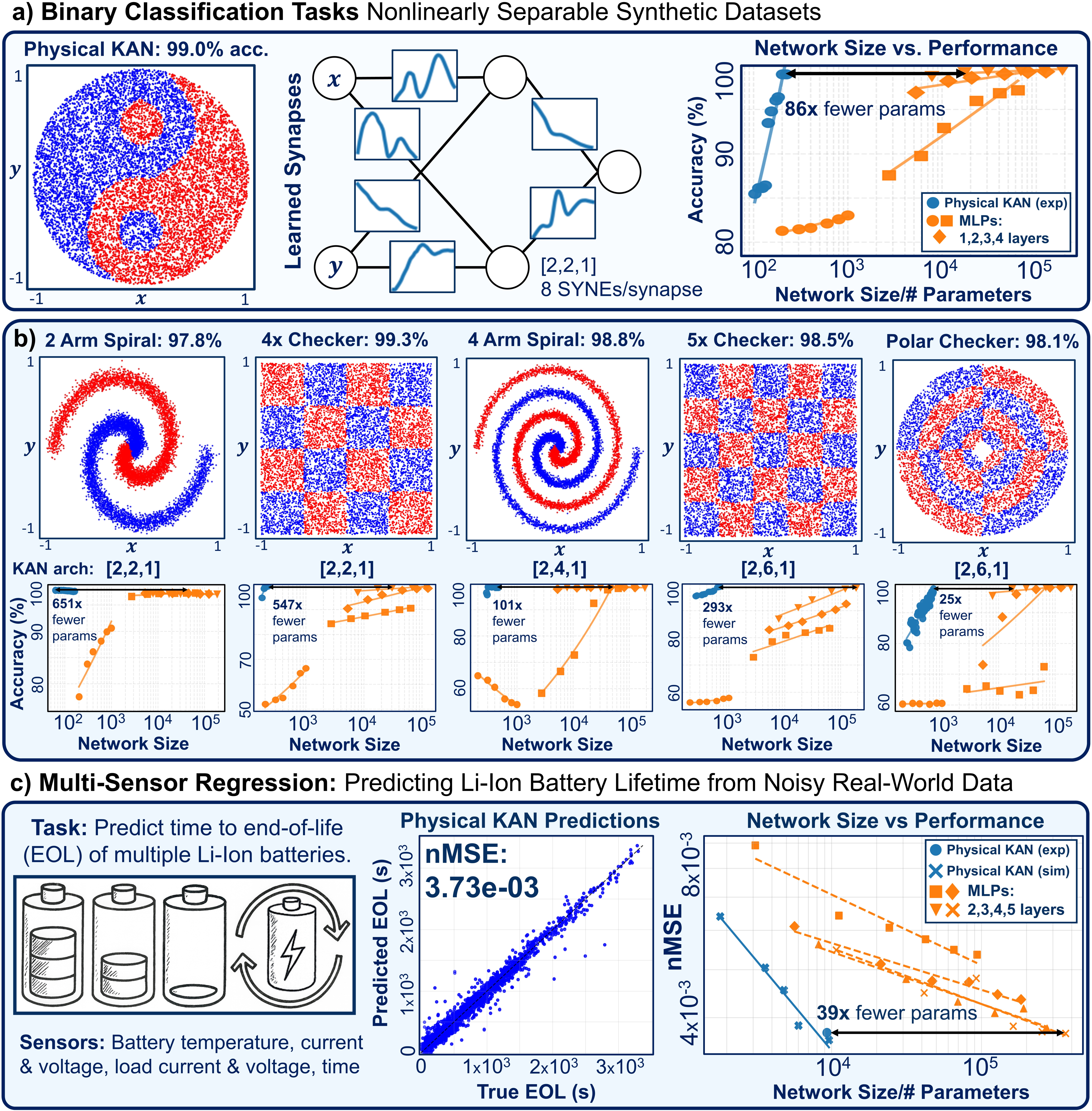}
\caption{\textbf{Binary Classification and Prediction of Li-Ion Battery Dynamics on Real-World Multi-Sensor Data.} \textbf{a)} Binary classification on nonlinearly-separable synthetic datasets. Here, physical KANs are trained on a set of 8k $x,y$ points ($x,y: -1,1$) which belong to one of two classes `Red' or `Blue', sampled from a variety of nonlinearly separable distribution shapes. The learned KAN parameters/synaptic shapes are then transferred to physical SYNE hardware, and performance evaluated on a test set of 10k points. The top-left panel shows experimental test performance on a `yin-yang' dataset with 99.0\% accuracy using a small [2,2,1] KAN with 8 SYNE devices per synapse. Learned synaptic shapes are shown on a schematic of the network. Classification performance vs. network size is evaluated for a range of physical KAN architectures relative to multilayer perceptrons of 1-4 hidden layers. Physical KANs outperform equivalently parameterised multilayer perceptrons, with up to 90$\times$ reduction in trainable parameter count relative to equally performing multilayer perceptrons. \textbf{b)} Five classification tasks with increasing difficulty left-to-right. Physical KANs outperform equivalently-parameterised multilayer perceptrons - with over 2 orders of magnitude reduction in trainable parameter count relative to multilayer perceptrons for four tasks. The 2-arm spiral accuracy is dominated by points located close to the fuzzy boundary between red and blue arms at the spiral centre, hence the lower accuracy even though the overall task and decision boundary is easier relative to other tasks. \textbf{c)} Prediction of Li-Ion battery dynamics and time to end of battery life (EOL) from noisy real-world multi-sensor data (NASA battery dataset). The other tasks considered in this study are ideal numerical functions or synthetically generated datasets, where the function-approximation benefits of the KAN are well suited. Here, we examine physical KAN performance on a real-world multi-sensor dataset recorded from noisy physical systems, a large set of Li-Ion batteries undergoing multiple charge/discharge cycles. The underlying charge dynamics can be described by physical equations, but the real-world behaviour of battery devices is non-ideal with experimental noise, differences between individual batteries, sensor and environmental drift, and other non-idealities. The task is to predict the time to end of life from sensor data, with no prior knowledge of how many charge/discharge cycles the battery has undergone. Here, physical KAN networks ([9,12,1], 12 SYNEs/synapse) outperform equivalently-parameterised multilayer perceptrons, demonstrating that SYNE-based physical KANs are not restricted to performing well on synthetic or numerically ideal tasks/datasets. The central plot shows experimental performance on the test set, plotting KAN predicted vs. true time to end of battery life over 547 discharge cycles. We compare performance vs. network size for physical KANs against multilayer perceptrons, showing that physical KANs are able to perform well at compact network sizes on real-world tasks, requiring 39$\times$ fewer parameters than equally-performing MLPs.
}
\label{Fig5} 
\end{figure}

\subsection*{Binary Classification}

To further explore the capability of SYNE-based physical KANs, we now explore their ability to classify nonlinearly-separable binary data. The process is essentially the same as the multivariate nonlinear function regression in Figure \ref{Fig4}, but here the smoothly varying functions are replaced by a range of two-class points (representing 0 or 1) with sharp transitions and nested decision boundaries. Here, the output neuron activation of the physical KAN is thresholded to provide a classification response, e.g. 0-0.5 = class A, 0.5-1 = class B. While KANs are known to be well suited to function regression tasks due to their smooth and tunable nonlinear basis functions and potential inductive biases from matching network architecture to function form, nonlinearly-separable classification tasks are a classic machine learning benchmark problem where many network architectures can perform well.

To demonstrate the process, we use a `yin-yang' shaped dataset shown in Figure \ref{Fig5}a). We sample 8,000 points to form a training set, and a further separate non-overlapping 10,000 points to form a test set. We train using backpropagation to minimise binary cross-entropy loss, then evaluate on the test set. The yin-yang plot in Figure \ref{Fig4}b) shows the experimental test set results and learned nonlinear synaptic functions, with a [2,2,1] physical KAN with 8 SYNEs per synapse achieving 99.0\% classification accuracy. We evaluate performance vs. network size of physical KANs with iteratively pruned SYNE device counts against 1-4 layer MLPs, and find that the best performing physical KAN requires 86$\times$ fewer trainable parameters than an equivalently performing MLP.

We assess classification performance against a set of five progressively harder classification tasks, increasing the number and complexity of nested class decision boundaries, shown in Figure \ref{Fig5}b). In each case, physical KANs outperform equivalently parameterised MLPs and in four of the five cases require over two orders of magnitude fewer parameters than equally capable MLPs, from 101$\times$ - 651$\times$ across the considered tasks. This demonstrates that physical KANs are not restricted to function approximators or smooth regression tasks.
We note that physical KANs outperform MLP networks more substantially on these classification tasks than the regression tasks presented earlier, with greater reductions in network size between physical KANs and equivalently-performing MLPs. This is likely due to the detrimental effects of experimental noise and model-reality gaps being suppressed by the thresholding between classes, only playing a substantial role in precision close to decision boundaries.


\subsection*{Real-world experimental datasets - Li-Ion Battery Aging - NASA Dataset}

So far, all tasks and datasets considered have been synthetic - numerically or mathematically generated, rather than experimentally measured real-world data. Good performance on synthetic data is a good indicator of network capability, but the capability of a machine learning system to learn well from non-ideal, real-world, experimentally measured data is a decisive factor in assessing whether emerging networks and physical devices can play a useful real-world role. Experimental noise, imperfect measurements and sensor drift typically make generalisation more challenging in these tasks than in idealised synthetic datasets. 

Here, we assess the ability of SYNE-based physical KAN networks to predict the time to end-of-life (EOL) of Li-Ion batteries from real-world multi-sensor data (NASA battery degradation dataset\cite{NASA_LiIonBatteryAging_2025}). An attractive property of physical computing networks is the ability to compute `in domain' with the network processing information in the modality of the real-world experimental input (e.g. processing optical signals with photonic neural networks), such that transduction between electrical/optical/digital regimes etc. can be avoided alongside associated energy and time costs. Here, we have selected a task where the multiple battery sensors output voltages, which can natively be input into our SYNE-based electronic physical KAN. While running software neural networks on digital hardware may consume enough energy that battery-monitoring edge-computing use-cases are unattractive, the reduced network sizes and low energy consumption of SYNE-based KANs may offer potential in such roles. 

The dataset contains 33 separate batteries with a distribution of charging dynamics and lifetimes, which are charged and then discharged under load over 2766 charge/discharge cycles. Batteries are attached to a range of experimental sensors: battery temperature, battery current, battery voltage, load current, load voltage, and time since a given discharge cycle began. The task is to predict the time between a given instantaneous measurement and the point when a minimum threshold value of remaining battery charge is reached, here 1.4 Ah from a total initial capacity of 2 Ah. Each battery has slightly different dynamics, which degrade over repeated charge/discharge cycles - no knowledge of which battery is under test or how many prior charge/discharge cycles have been performed is provided to the networks in the task. We sample 50 measurements from each discharge cycle, giving 138,300 samples which are split 70/10/20\% into train, validation, and test sets. 

We train a physical KAN network with each of the sensor values feeding an input neuron, and a single output neuron which outputs EOL predictions. The network has a hidden layer of 12 neurons and all synapses consist of 12 SYNEs. The `Physical KAN predictions' panel in Figure \ref{Fig5}c) shows the test set results comparing true EOL (x-axis) against physical KAN prediction (y-axis). Perfect performance looks like a straight $y = x$ line, with any deviation from this an error/inaccuracy in prediction. The physical KAN network is able to learn the underlying dynamics, with a clear trend around $y = x$ and nMSE value of 3.73e-3. We compare network size vs. performance between 2-5 layer MLPs and physical KANs (a single experimental network and multiple digital twin simulated networks with iteratively pruned SYNE device counts), and find that our experimentally measured physical KAN outperforms equivalently parameterised MLPs, with MLPs requiring 39$\times$ more parameters to match performance. This is a significant result, showing that on real-world data with no architectural inductive biases from attempting to match the KAN topology against a known function structure, physical KANs are able to learn the underlying nonlinear dynamics governing complex physical systems (the Li-Ion batteries) and perform useful tasks. Using circuit-level estimates for our readout and control electronics (methods), we project 135 nJ scale per-inference energy for the battery models studied here, versus 15-180\,$\upmu$J-scale costs for equivalently-performing multilayer perceptron baselines on conventional CPU/GPU hardware (methods). These considerations make multi-sensor battery end-of-life prediction and related edge-computing tasks integrating multiple voltage-based physical sensors a potentially promising application domain for physical KAN hardware: a real-world, noisy regression problem where compact, efficient networks at the edge can provide benefits.

\section*{Conclusions}

We have experimentally demonstrated a physical implementation of a Kolmogorov-Arnold Networks (KAN) by learning synaptic nonlinearities directly in voltage-programmable silicon-on-insulator Synaptic Nonlinear Element (SYNE) devices. This work demonstrates a practical route to reducing the required device count and network size relative to linear-weight based networks, and is able to outperform parameter-matched software MLP baselines despite experimental noise and model-reality gaps between physical devices and the differentiable digital twin employed for training. We show that KAN architectures are able to exploit rich, locally reconfigurable heterogeneous nonlinear physical device dynamics, shifting the computational burden away from large arrays of identical linear weights, towards expressive and programmable nonlinear physics which are a key strength of computing in physical networks. SYNE devices operate at room temperature, 2 MHz throughput, without observed degradation over months-long experimental runs, and are fabricated using mature SOI semiconductor processing - compatible with on-chip or chiplet based integration\cite{raskin_SOI_2022}.

Across a range of function regression, classification, and real-world multi-sensor Li-Ion battery EOL prediction tasks, we show that physical KANs are able to outperform equivalently parameterised perceptron-style networks based on training linear weights including on noisy experimentally-measured datasets. We show reductions in trainable parameter count (and device count for physical networks) of up to two orders of magnitude relative to equivalently-performing MLPs, promising for tackling scaleability issues facing physical neural networks. A central result is the demonstration that while a nonlinear physical device may not provide the arbitrary function representation of a digital spline, combining a small number of devices in series and parallel leads to expressive physical synapses which can outperform linear-weight network baselines despite experimental noise. We introduce an $\epsilon$-packing based approach `Epsilon Expressivity' to quantify the nonlinear expressivity of devices or networks, which closely tracks function regression performance and can be used as a compute-efficient design tool for arbitrary systems. We believe these results can serve as inspiration that the gains observed here when transitioning learning to physical nonlinearities may be implemented in a wide range of physical systems and devices.

It is worth noting that we have arrived at architectures based on training linear weight matrices dominating neural networks not because these are universally the most powerful way of implementing computational networks, but because they are implemented very efficiently as matrix multiplications on GPU hardware. This co-evolution of learning architectures following the strengths and weaknesses of the dominant type of computational hardware has been referred to as `the hardware lottery'\cite{hooker2021hardware}, `the bitter lesson'\cite{sutton2019bitter}, and stated as `the hardware \textit{is} the software'\cite{laydevant2024hardware}. As new physical systems emerge with promising nonlinear dynamics, intrinsic memory, and appealing energy efficiency, it is unlikely that existing GPU-optimized linear weight based networks are the optimal architecture - and indeed likely neither will be `vanilla' KANs as described so far, but the results shown here demonstrate that new architectures which refocus learning and computation into areas more directly synergistic with emerging device physics will be crucial to explore.

Looking ahead, future iterations of this approach will require implementations that combine arrays of devices with on-chip gain, weighting and summation, and non-volatile parameter storage, integrating the linear operations which are currently performed digitally off-chip. In parallel, while backpropagation through differentiable digital twins provides an effective training route in the present work, future systems will benefit from physics-aware learning strategies that reduce dependence on backward passes and detailed differentiable models, including local learning rules, noise-aware training\cite{manneschi2024noiseawaretrainingneuromorphicdynamic} and forward-only training approaches\cite{momeni2025training,chen2025self}.

Together, our findings support the idea that alongside refining physical devices\cite{markovic2020physics,mcmahon2023physics,kudithipudi2025neuromorphic} and physics-compatible training algorithms\cite{momeni2025training}, developing network architectures capable of learning and harnessing the rich and varied heterogeneous nonlinear dynamics available in physical systems can play a valuable role in progressing the field, reducing network size, device count, and energy by focusing computation onto where physical neural networks can provide unique benefits against conventional hardware platforms.

\subsection*{Author contributions}

\textbf{Fabiana Taglietti} co-designed the Silicon-on-Insulator SYNE devices, co-proposed the implementation of Kolmogorov-Arnold Networks in Silicon-on-Insulator SYNE devices, performed the electrical transport measurements and comprehensive electrical characterization, co-developed the QM OPX+ FPGA code for the electrical measurements, collected the dataset, prepared figures, contributed to writing, reviewing and editing the manuscript. 

\textbf{Andrea Pulici} designed the chip with Silicon-on-Insulator SYNE devices, optimized the doping process for the doping of the silicon on insulator substrates, performed estensive electrical characterization of the doped SOI substrates, designed and optimized the lithography process flow for the fabrication of the SYNE devices, fabricated the Silicon-on-Insulator SYNE devices, contributed to the analysis of the electrical measurements, prepared figures, contributed to writing, reviewing and editing the manuscript

\textbf{Maxwell Roxburgh} performed MLP benchmarks for the multivariate nonlinear function regression and classification tasks, developed the iterative pruning approach for reducing network size by removing SYNE devices from specific synapses, performed analysis quantifying the difficulty of multivariate function regression tasks, contributed to analysis of expressivity, contributed to the development, training and evaluation of the digital twin and machine learning framework, and provided review and edits of the manuscript.

\textbf{Gabriele Seguini} co-optimized the process for the doping of the silicon on insulator substrates, co-designed the Silicon-on-Insulator SYNE devices, contributed to electrical characterization of the doped SOI substrates, co-designed and contributed to optimized the lithography process flow for the fabrication of the SYNE devices, prepared figures, contributed to writing, reviewing and editing the manuscript.

\textbf{Ian Vidamour} contributed to the design and implementation of the machine learning framework, contributed to machine learning code and analysis, provided key feedback and revisions on figures and text, and co-design of the physical Kolmogorov-Arnold Network concept.

\textbf{Stephan Menzel} contributed to the energy consumption analysis, developed the VCM memristor scheme for tuning voltages, writing, reviewing and editing the manuscript.

\textbf{Edoardo Franco} co-developed the QM OPX+ FPGA code for the electrical measurements.

\textbf{Eleni Vasilaki} contributed to the design of the machine learning framework, contributed to machine learning code and analysis, provided key feedback and revisions on figures and text, valuable discussions on the scope, direction and narrative of the study, and co-design of the physical Kolmogorov-Arnold Network concept. Eleni obtained funding covering some aspects of the project, and proposed the Li-Ion battery dynamics task.

\textbf{Michele Laus} designed and synthesized the P terminated polymers, performed characterization of the P terminated polymers, co-design the precision polymer doping protocol, co-optimized the process for the doping of the Silicon-on-Insulator substrates, critical reading and  reviewing  the manuscript.

\textbf{Michele Perego} designed the precision polymer doping protocol, co-optimized the process for the doping of the silicon on insulator substrates, co-designed the Silicon-on-Insulator SYNE devices, co-proposed the implementation of Kolmogorov-Arnold Networks in Silicon-on-Insulator SYNE devices, co-designed the lithography process flow for the fabrication of the SYNE devices, prepared figures, contributed to writing, reviewing and editing the manuscript

\textbf{Thomas J. Hayward} co-proposed the implementation of Kolmogorov-Arnold Networks in Silicon-on-Insulator SYNE devices, lead the development of the digital twin training and machine learning framework, performed machine learning experiments, proposed the use of epsilon-packing as a quantitative metric of expressivity, performed the energy efficiency analysis and proposed future circuit design, provided feedback and editing throughout figure and manuscript development.

\textbf{Marco Fanciulli} conceived and co-design the Silicon-on-Insulator SYNE devices and the use of the QM OPX+ for the high-throughput FPGA controlled transport measurements, co-proposed the implementation of Kolmogorov-Arnold Networks in Silicon-on-Insulator SYNE devices, coordinated the SYNE electrical measurements and their analysis, contributed to writing, reviewing and editing the manuscript.

\textbf{Jack C. Gartside} co-proposed the implementation of Kolmogorov-Arnold Networks in Silicon-on-Insulator SYNE devices, performed electrical measurements and analysis and processing for the KAN experiments, co-developed FPGA code for the electrical measurements, co-developed the machine learning framework, performed MLP benchmarks for the univariate function regression and Li-Ion battery dynamics tasks, developed initialisation schemes for the physical KAN parameters, co-developed expressivity metrics and performed expressivity analysis, made figures, wrote the manuscript, and obtained funding for aspects of the project.

\subsection*{Acknowledgements}
JCG was supported by the Royal Academy of Engineering Research Fellowship RF2122-21-363. \\
JCG was supported by the EPSRC ECR International Collaboration Grant EP/Y003276/1.\\
JCG was supported by the ERC Starting Grant MORPHON.\\
JCG was supported by the Imperial College London President's Excellence Fund for Frontier Research.\\
Maxwell Roxburgh was supported by the Imperial College London Val O'Donoghue PhD Fellowship.\\
EV was supported by MARCH EP/V006339/1, ActiveAI EP/S030964/1, Causal XRL EP/V055720/1, and the EPSRC Neuroware IKC. \\ 
TJH was supported by EPSRC grants: MARCH EP/V006339/1, the EPSRC Neuroware IKC and Spintronic Reservoir Fusion. \\ 
IV was supported by MARCH EP/V006339/1 and the EPSRC Neuroware IKC.\\
MF, MP and GS were supported by the PRIN project "Dopants Networks in Silicon for Unconventional Computing in Materia - DONORS" financed by the European Union - Next Generation EU, Mission 4 Component 1 CUP 2022WBPHKF.\\
FT, MF, and JCG acknowledge Dr. Valerio Di Palma for support during the experimental phase of this study. \\
FT, EF, MF, and JCG acknowledge the Quantum Machine support team for helpful assistance and discussion during the optimization of their setup.\\
FT and MF acknowledge Prof. W.G. van der Wiel and his group at the University of Twente for valuable discussions regarding the DNPU concept.\\
JCG acknowledges Tristan da C\^amara Santa Clara Gomes for valuable discussion and feedback on the manuscript.\\


\subsection*{Competing interests}
The authors declare no competing interests.

\subsection*{Dual-use statement}
FT, AP, MR, GS, IV, EF, EV, MP, MF and JCG declare their commitment to responsible scientific conduct and explicitly oppose any dual-use or otherwise harmful application of the findings presented in this work.

\section*{Methods}

\subsection*{Experimental methods}

\subsection*{Fabrication}

Silicon-on-insulator `Synaptic Nonlinear Element' devices were fabricated on commercially available SOI wafers and packaged for room-temperature electrical characterisation. A $1\times1$~cm$^2$ chip was diced from a 70~nm-thick SOI wafer and cleaned sequentially in acetone and isopropanol ultrasonic baths. The SOI device layer was thinned to $H_{\mathrm{SOI}}\approx30$~nm by thermal oxidation at 1000~$^\circ$C followed by removal of the grown SiO$_2$ in HF solution; the process was optimised to avoid observable sample degradation (full process details are reported in Ref.~\cite{pulici_electrical_2023}).

\begin{figure}[t!]
\centering
\includegraphics[width=0.5\textwidth]{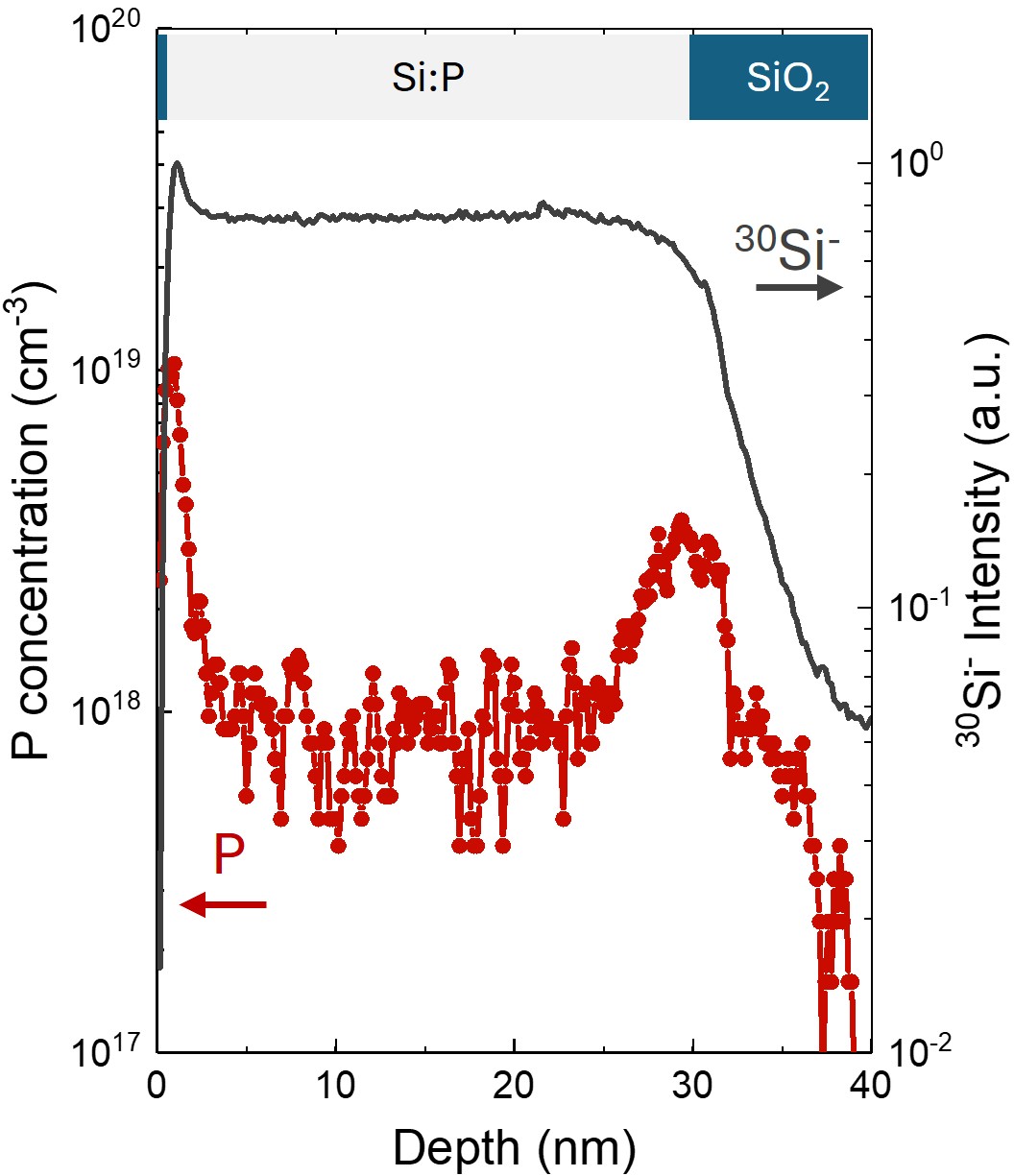}
\caption{\textbf{} 
Calibrated ToF-SIMS depth profile of the phosphorus distribution
into the silicon device layer, showing a uniform dopant concentration of $9 \times 10^{17} cm^{-3} $ throughout the entire layer depth with small P accumulation at the $Si/SiO_2$ interfaces. $^{30}Si^{-}$ profile is also reported for reference, highlighting the interface between silicon device layer and buried oxide, in the SOI structure.}
\label{Doping-profile} 
\end{figure}

Phosphorus doping of the ultrathin SOI device layer was performed using a  bottom-up approach based on polystyrene end-terminated with a P-containing moiety (PS--P), which are grafted on the SOI substrate to form a surface $\delta$-layer dopant source on the deglazed silicon surface\cite{perego_control_2018,pulici_electrical_2023,pulici_arxiv_2026}. To achieve accurate control of dopant dose and redistribution, the process was divided into two rapid thermal anneals (RTP) in N$_2$. The first anneal (1000~$^\circ$C for 1~s) sets the injected P dose. After removal of the polymer $\delta$-layer source, a second anneal (1000~$^\circ$C for 100~s) redistributes and activates the dopants to yield a uniform concentration through the device layer\cite{pulici_electrical_2023}. Calibrated ToF-SIMS measurements confirmed a uniform phosphorus profile, with an average dopant density of $9\times10^{17}$~cm$^{-3}$ (Fig.~\ref{Doping-profile}).

Following doping, a thin ($\sim$2~nm) surface SiO$_2$ layer was chemically grown by SC2 cleaning (H$_2$O:H$_2$O$_2$:HCl, 5:1:1) at 75~$^\circ$C for 20~min. Device regions were defined by photolithography and wet etching in 20~wt.\% KOH at room temperature to open 20~$\upmu$m-diameter windows. The surface was then treated with O$_2$ plasma to remove residual contaminants. Aluminium contacts (80~nm Al) were patterned by a second photolithography step and deposited by thermal evaporation to form eight electrodes and define a $\sim$6~$\upmu$m-diameter active region between contacts (Fig.~\ref{Fig1}). Finished chips were wire-bonded to a printed circuit board (PCB) and connected to SMA connectors for electrical measurements; the PCB was enclosed in a metal housing with external SMA access.

\subsection*{Electrical Transport Measurements} 
\subsubsection*{Initial Device Characterisation Transport Measurements} 

The initial electrical $I$--$V$ characterization of the SYNEs before performing the Kolmogorov-Arnold Network experiments was performed at room temperature with a Keithley 4200A-SCS (Semiconductor Characterization System) Parameter Analyzer, equipped with four 4200 SMU units with 4200-PA preamplifiers. 
In the configuration found to give the best range of nonlinear $I$--$V$ shapes, four electrodes were connected to the SMUs with the remaining electrodes kept floating. 
$I$--$V$ characteristics were measured by applying a voltage sweep from -2\,V to 2\,V at the input electrode, and measuring the output current at the opposite grounded electrode. To verify the absence of hysteresis, the input voltage is swept from 0 to 2\,V, then from 2\,V to -2\,V and back to 0\,V, with no observed hysteretic behaviours. A voltage step of 0.01\,V was employed and a sweep rate of 0.05\,V/s was determined by the fixed current measure range of 1\,mA. 
Constant tuning voltages $V_\mathrm{Tune}$ in the range [-3\,V,3\,V] were applied at the two remaining electrodes for the entire duration of the measurement to characterise the SYNEs' expressivity and range of accessible nonlinear $I$--$V$ shapes. Devices were operated at room temperature and without back-gating. Typical output currents in the operating range were $\sim$0.1-1~$\upmu$A.

\subsubsection*{High-throughput FPGA controlled transport measurements for KAN experiments}

To measure training data for the digital twin of our SYNE devices and carry out the experimental results for the Physical Kolmogorov-Arnold Networks, a measurement setup was developed based on a Quantum Machine Operator-X (OPX+), a fast MHz FPGA-based source measurement unit configured with 10 outputs and 2 inputs. The OPX+ performs fast output waveform generation to send voltage information to the SYNE device inputs (both input data and tuning voltages), and acquisition and processing of the device's output. For our experimental results, the system was operated at 2 MHz, e.g. $2\times10^6$ discrete datapoints per second were run through the SYNE device.

A transimpedance amplifier was used to connect the SYNE output to the measurement input of the OPX+ FPGA source-measurement unit, with a gain of  $10^5$ V/A. Tuning voltages learned by backpropagation through the differentiable digital twin were applied to the hardware device, after which input sequences corresponding to the target dataset were driven through the SYNE and $I_{\mathrm{Out}}$ recorded. Input sequences were ordered from negative to positive $V_\mathrm{In}$ values by the FPGA.

\subsection*{Computational methods}

\subsection*{Training the SYNE digital twin}

To model the SYNE device physics in a differentiable surrogate model, a data-driven digital twin neural network was trained on experimentally measured $I$--$V$ sweeps under a range of different control voltages. The digital twin is a multilayer perceptron of architecture [3,200,200,200,1] with ReLU activations on neurons, with the three inputs relating to $V_{In}, V_{Tune 1}, V_{Tune 2}$ and the output being $I_{Out}$. 
The digital twin is differentiable, allowing us to use backpropagation/gradient descent to determine which experimental tuning voltage values $V_{Tune}$ are needed to give nonlinear $I$--$V$ traces of a desired shape/curvature. 

The experimental data set used to train the digital twin was obtained as follows: Control voltages of $\pm$2\,V were used on each of the two tuning voltage $V_\mathrm{Tune}$ inputs, with 15 incremental measurements between -2\,V and +2\,V on each tuning voltage input, giving a total of $15\times15 = 225$ $I$--$V$ sweeps. For each sweep, the input voltage was swept from -2\,V to +2\,V in 10k steps while measuring the output current. Twenty percent of the data points were held back during training to form an unseen test set, with a test MSE of 4.4e-7 for the digital twin on the experimental $I$--$V$ test set. Training of the digital twin was done via backpropagation using ADAM optimiser, 3000 training epochs, a batch size of 8192, and learning rate of 1e-4.

SI Figure \ref{SI-Digi-Twin-Test} shows 100 representative experimentally measured $I$--$V$ traces (solid lines), and the digital twin emulation for the same input parameters (dashed lines), showing good correspondence and the ability of the digital twin to learn the voltage-controlled nonlinear $I$--$V$ dynamics of SYNE devices.

\subsection*{Training the digital twin based SYNE Kolmogorov-Arnold Network}

Network parameters ($V_\mathrm{Tune}$, output gains, etc) were initialised according to a Xavier scheme\cite{glorot2010understanding,rigas2025initialization} then trained using backpropagation and gradient descent via the differentiable data-driven digital twin with Adam optimiser and a learning rate of 1e-4. Penalty terms are imposed to keep output gain values below 3$\times$ and restrict $V_\mathrm{Tune}$ values from entering ranges where experimental signal-to-noise becomes poor (below -0.6\,V), giving a useful range of $V_\mathrm{Tune}$ = -0.6\,V to 2\,V. We also imposed penalty terms to keep the digital twin operating in parameter ranges seen during training on experimental $I$--$V$ data, to reduce model-reality gap and keep the learned parameters to ranges where the SYNE device is known to operate well. Xavier initialisation gives substantially better results than purely random parameter initialisation (Fig. \ref{SI-Random-vs-Xav}) as observed previously in software KANs\cite{rigas2025initialization}. 

After transferring the learned parameters to the experimental SYNE device, performance was evaluated on a validation set of points and the linear parameters (linear scaling of SYNE responses and neuron biases) adjusted to minimise loss on the validation set. These updated parameters were saved, and used to perform the test set evaluations.

\subsection*{Estimating power consumption}

To estimate the energy per inference of physical Kolmogorov--Arnold Networks, we calculate the power consumed by both a SYNE device and the additional analogue electronic components required to realise a non-volatile network edge. We envisage that memristors based on valence-change memory (VCM) devices\cite{dittmann2021nanoionic} are placed in series with the SYNE input contact and with all tuning contacts. Assuming that the lowest-resistance state of the VCM is comparable to that of the SYNE device, and that the highest-resistance state is orders of magnitude larger, the VCM acts as a programmable voltage divider. This enables non-volatile storage of (a) the effective input-range scaling and (b) the amplitudes of the tuning voltages. Output amplification is assumed to be provided by a low-power operational amplifier. A further VCM device is incorporated into the feedback path of the amplifier, where it acts as a programmable feedback resistance. In this configuration, the non-volatile resistance state of the VCM directly sets the closed-loop gain of the amplifier, allowing the output gain of the network edge to be tuned and retained without static power consumption. While we restrict the gains/scaling to a maximum of $3\times$ during network training, we perform the projected calculations below using a gain of $5\times$ to allow some tolerance and additional headroom.

\paragraph{SYNE power consumption.}
We estimate the resistance of the SYNE device by measuring its current--voltage characteristics with the output contact and both tuning contacts connected to ground. In this configuration, symmetry of the device geometry implies that the current injected at the input contact is distributed equally between the two tuning contacts, such that the current measured at the output contact is representative of the current flowing through each branch of the device. From these measurements, we extract an effective device resistance of $1.1~\mathrm{M}\Omega$. During normal operation, the SYNE device experiences potential differences of order $1~\mathrm{V}$. At this operating point, the corresponding power dissipation is therefore approximately $0.9~\upmu\mathrm{W}$ per SYNE.

\paragraph{VCM power consumption.}
When the VCM devices are used as programmable voltage dividers, they are biased from supply rails at $\pm 2\,V_{\max}$ in order to allow tuning voltages up to $\pm V_{\max}$ to be applied at the tuning contacts when the VCM is in its lowest-resistance state. In our measurements, $V_{\max}=2~\mathrm{V}$, requiring supply rails of $\pm 4~\mathrm{V}$. For a typical tuning voltage of $\sim 1~\mathrm{V}$, the VCM resistance must be approximately three times larger than that of the SYNE device to realise the required voltage division. Under these conditions, the resulting current through the VCM--SYNE series pair leads to a power dissipation of approximately $2.7~\upmu\mathrm{W}$ per VCM.

\paragraph{Amplifier power consumption.}
While dedicated CMOS ASIC implementations may ultimately offer lower power consumption and co-integration of amplifiers and SYNEs on SOI platform chips/chiplets, here we consider the use of standard off-the-shelf integrated-circuit operational amplifiers to provide a baseline. For our estimates we take as an example the OPA357 operational amplifier from Texas Instruments, which provides a gain--bandwidth product of $100~\mathrm{MHz}$ at quiescent currents of order milliamps and is suitable for transimpedance configurations.\cite{ti_opa357_ds} We assume a maximum required closed-loop gain of $5$, giving an available closed-loop bandwidth of $\sim 20~\mathrm{MHz}$. In the calculations below we fix the operating frequency to $2~\mathrm{MHz}$ (corresponding to operation at $\sim 1/10$ of the available closed-loop bandwidth, providing settling margin). Since the currents flowing through the SYNE devices are small, we assume that the amplifier power consumption is dominated by the quiescent current rather than by output drive. At a supply voltage of $4~\mathrm{V}$, the quiescent current of $4.9~\mathrm{mA}$ therefore results in a power dissipation of approximately $20~\mathrm{mW}$ per amplifier.

\paragraph{Amplifier location within network topology.}
The simplest case is where each SYNE device has its own amplifier (`amplifier-per-SYNE'). We also consider `amplifier-per-neuron' schemes, where each SYNE output current is first relatively weighted by a VCM memristor and then Kirchhoff-summed, with amplification applied only once per neuron.

\paragraph{Projected network energy per inference: 5$\times$ Checker task.}
We estimate the energy required to perform a single inference operation on the classification task `5$\times$ Checker' shown in Fig.~\ref{Fig4}b. The network has an architecture of $[2,6,1]$, with eight SYNE devices used in parallel to realise each network edge, resulting in an initial network size of $144$ SYNE devices. Following network pruning, this was reduced to $124$ SYNE devices without loss of classification performance, and all energy estimates are based on this pruned architecture. We fix the operating frequency to $f=2~\mathrm{MHz}$, corresponding to an inference time $T=1/f=0.5~\upmu\mathrm{s}$.

\emph{Amplifier-per-SYNE} Each SYNE requires three VCM units (one for input scaling and two for tuning-voltage inputs) and one amplifier. The per-SYNE power is
$P_{\mathrm{SYNE}} \approx P_{\mathrm{amp}} + (0.9 + 3\times 2.7)\,\upmu\mathrm{W} \approx 20~\mathrm{mW} + 9.0~\upmu\mathrm{W}$,
so the per-SYNE energy per input at $2~\mathrm{MHz}$ is
$E_{\mathrm{SYNE}} \approx P_{\mathrm{SYNE}}T \approx (20~\mathrm{mW})\times(0.5~\upmu\mathrm{s}) \approx 10~\mathrm{nJ}$.
For $124$ SYNE devices, this yields a total network energy of approximately
$E_{\mathrm{net}} \approx 124\times 10~\mathrm{nJ} \approx 1.2~\upmu\mathrm{J}$ per inference.

\emph{Amplifier per-neuron} If SYNE outputs are first relatively weighted by an additional output VCM per SYNE and then Kirchhoff-summed, amplification can be applied once per neuron. For the $[2,6,1]$ classifier (18 edges, 7 post-synaptic neurons excluding inputs), the total power is
$P_{\mathrm{net}} \approx N_{\mathrm{amp}}P_{\mathrm{amp}} + 124\,(0.9 + 4\times 2.7)\,\upmu\mathrm{W}$.
This gives at $2~\mathrm{MHz}$:
(i) amplifier-per-edge ($N_{\mathrm{amp}}=18$): $E_{\mathrm{net}}\approx \big(18\times 20~\mathrm{mW}\big)T \approx 180~\mathrm{nJ}$;
(ii) amplifier-per-neuron ($N_{\mathrm{amp}}=7$): $E_{\mathrm{net}}\approx \big(7\times 20~\mathrm{mW}\big)T \approx 70~\mathrm{nJ}$.
In all cases the total is dominated by amplifier quiescent power; amplifier sharing therefore provides a direct route to reducing energy without changing the SYNE device itself.

\paragraph{Projected network energy per inference: Li-Ion battery EOL prediction task}
For the battery task we consider a physical KAN with architecture $[9,12,1]$ using 12 SYNE devices in parallel per synapse. The number of edges is $9\times 12 + 12\times 1 = 120$, giving a total of $120\times 12 = 1440$ SYNE devices. At $2~\mathrm{MHz}$:
(i) amplifier-per-SYNE gives $E_{\mathrm{net}}\approx 1440\times 10~\mathrm{nJ}\approx 14.4~\upmu\mathrm{J}$ per inference;
(ii) amplifier-per-neuron uses $N_{\mathrm{amp}}=12+1=13$ amplifiers and gives
$E_{\mathrm{net}}\approx (13\times 20~\mathrm{mW})T \approx 130~\mathrm{nJ}$ per inference (with SYNE+VCM conduction a small correction at this operating point).

For a comparably performing software linear-weight network baseline, an MLP with architecture $[5,300,300,300,300,300,1]$ contains $\sim 3.61\times 10^5$ parameters (Fig. \ref{Fig5}).

\paragraph{Comparison to Software MLPs Implemented on Digital CMOS Hardware.}
We estimate the energy per inference required for a conventional multilayer perceptron (MLP) implemented on an NVIDIA RTX~4090 GPU to match the performance of the hardware KANs described above. An MLP with architecture $[2,300,300,300,1]$ was found to be necessary to achieve comparable classification performance on the \textbf{5$\times$ checker task}. Evaluating this network requires approximately $1.8\times 10^{5}$ multiply--accumulate operations, corresponding to a total of $\sim 3.6\times 10^{5}$ floating-point operations (FLOPs). The RTX~4090 has a peak FP32 throughput of $82~\mathrm{TFLOPs/s}$ and a typical power draw of $\sim 450~\mathrm{W}$ under load. Under the idealised assumption of full utilisation, this yields an energy per inference of approximately $2~\upmu\mathrm{J}$, around six times larger than that projected for the 124-device physical KAN and approximately 40 and 100 times larger than the amplifier-per-edge and amplifier-per-neuron schemes. However, this estimate (i) assumes perfect utilisation of the GPU's computational resources and (ii) neglects latency associated with kernel initialisation, scheduling, and memory access, which dominate the execution time for small networks. To obtain a more realistic estimate of the inference energy, we therefore measured the wall-clock time required for a single-datum forward pass through the network using the PyTorch framework, with an inference batch of 10,000 data points per trial, and averaged runtime per data point measured over 500 repetitions to be $5.1~\mathrm{MHz}$. We measured an average inference latency of approximately $0.2~\upmu\mathrm{s}$. Assuming a power draw of $450~\mathrm{W}$, this corresponds to an energy per inference of approximately $90~\upmu\mathrm{J}$, which is two to three orders of magnitude larger than the estimated energy consumption of the physical KAN implementations.

As an additional point of reference for realistic edge deployment, we consider the NVIDIA Jetson Nano, which operates in $5$--$10~\mathrm{W}$ power modes and provides a peak FP32 throughput of $\sim 2.36\times 10^{11}~\mathrm{FLOPs\,s^{-1}}$ (order-of-$10^{2}~\mathrm{GFLOPs\,s^{-1}}$). For the $[2,300,300,300,1]$ MLP above ($\sim 3.6\times 10^{5}$ FLOPs per inference), this implies an optimistic lower bound of $\sim 4$--$15~\mu\mathrm{J}$ per inference under full utilisation. In practice, for small networks and batch-1 operation, utilisation and software overhead typically dominate, so realised inference energies are expected to be higher than this bound (often in the $\sim 10$--$10^{2}~\mu\mathrm{J}$ range), remaining two to four orders of magnitude above the projected nJ-class physical-KAN implementations.

For the \textbf{Li-Ion battery EOL prediction task}, an MLP with comparable performance to the physical KAN has architecture $[5,300,300,300,300,300,1]$. This requires $\sim 3.6\times 10^{5}$ multiply--accumulate operations (weights) per inference, corresponding to $\sim 7.2\times 10^{5}$ FLOPs under the same $2\times$ (multiply+add) convention used above. Using the RTX~4090 peak FP32 throughput of $82~\mathrm{TFLOPs/s}$ and a representative power draw of $\sim 450~\mathrm{W}$, an optimistic full-utilisation lower bound is therefore $\sim 4~\upmu\mathrm{J}$ per inference. As with the smaller $[2,300,300,300,1]$ network, this bound substantially underestimates realised energy for batch-1 or small-batch inference where kernel launch, scheduling, and memory traffic dominate. Using the measured wall-clock latency of $\sim 0.2~\upmu\mathrm{s}$ for the $[2,300,300,300,1]$ network as a reference point, and noting that the Li-Ion MLP has $\sim 2\times$ higher FLOP count, a simple FLOP-proportional extrapolation suggests a latency of order $\sim 0.4~\upmu\mathrm{s}$ and hence an energy per inference of order $(450~\mathrm{W})\times(0.4~\upmu\mathrm{s})\sim 180~\upmu\mathrm{J}$ on the RTX~4090, with the same caveat that software/IO overhead can dominate and introduce additional task- and implementation-dependent variation.

Applying the corresponding optimistic throughput bound to the NVIDIA Jetson Nano (peak FP32 throughput $\sim 2.36\times 10^{11}~\mathrm{FLOPs,s^{-1}}$ and $5$--$10~\mathrm{W}$ operating modes), the $[5,300,300,300,300,300,1]$ network implies a best-case lower bound of $\sim 15$--$30~\upmu\mathrm{J}$ per inference under full utilisation. In practice, for small networks and batch-1 operation, utilisation and framework overhead typically dominate on embedded platforms as well, so realised inference energies are expected to exceed this bound (often by orders of magnitude depending on the software stack), remaining far above the projected nJ-class physical-KAN implementations discussed above.

\paragraph{Higher Efficiency Amplifiers: Switched-Capacitor Schemes (projection).}
As an alternative to continuously biased amplification, scaling can be implemented using a switched-capacitor charge-integration stage, which converts the Kirchhoff-summed synaptic current to a voltage by integrating onto an effective capacitance over a fixed window $T$ (current-in/voltage-out by integration).\cite{dei2024slew} In this approach, relative synaptic weighting is still provided by the VCM devices, while the neuron gain (current-to-voltage transimpedance) is set by the integration time and capacitance, $Z_{\mathrm{eq}}\approx T/C_{\mathrm{int}}$. For example, mapping $1~\upmu\mathrm{A}$ summed current to $\sim 1~\mathrm{V}$ at $2~\mathrm{MHz}$ corresponds to $Z_{\mathrm{eq}}\sim 1~\mathrm{M}\Omega$ and thus $C_{\mathrm{int}}\sim T/Z_{\mathrm{eq}}\sim 0.5~\mathrm{pF}$. The dominant energy then becomes the SYNE and VCM conduction during the compute window:
for the pruned $[2,6,1]$ network with output weighting (4 VCM per SYNE), $P_{\mathrm{cond}}=124(0.9+4\times 2.7)\,\upmu\mathrm{W}=1.45~\mathrm{mW}$, giving
$E_{\mathrm{cond}}=P_{\mathrm{cond}}T\approx 0.73~\mathrm{nJ}$ at $2~\mathrm{MHz}$.
The additional switched-capacitor switching/reset overhead scales as $\sim N_{\mathrm{neu}}\,C_{\mathrm{int}}V^2$ and is typically pJ-class for sub-pF to few-pF capacitors, making it a small correction at this operating point. Switched-capacitor circuits are standard CMOS mixed-signal building blocks and have been demonstrated in SOI technologies, with potential for future network designs exploring co-integration or co-packaging of SOI-compatible amplification alongside SYNE devices.

\subsection*{Quantifying Expressivity}

Several scalar metrics have been used to characterise physical computing substrates\cite{kurebayashi2025technical,grollier2020neuromorphic,markovic2020physics}, particularly in reservoir computing\cite{love2023spatial}, including measures of nonlinearity, kernel rank/separation, and information processing capacity, but these do not directly describe the expressivity, or the breadth/diversity of non-degenerate nonlinear functions accessible by a device or network. Existing metrics from the reservoir computing community are typically defined for a \emph{fixed dynamical reservoir} under a specified driving distribution and decoded by a chosen readout (often linear), and therefore do not directly quantify the design question we face here: the \emph{diversity of a programmable family of static synaptic transfer functions}. ``Nonlinearity'' scores capture deviation from linearity but not the number of meaningfully distinct nonlinear shapes available under tuning; kernel-rank style measures assume high-dimensional state trajectories and saturate or become noise-dominated for low-dimensional, curve-based primitives; and capacity measures such as IPC are computationally expensive and depend on task basis, input statistics and readout regularisation. During this project, we attempted to implement and compute a range of metric approaches, including Shannon entropy and cosine similarity, but were not able to observe clear trends or direct correlations with observed network performance. 

There is a need for easy to implement, computationally cheap, task agnostic metrics for quantifying nonlinear expressivity - hence we adopt and propose an epsilon-packing approach based on embedding the nonlinear curves accessible by a device/network as a point cloud in PCA space and calculating packing numbers to directly measure the diversity of achievable transfer-curve shapes, which we term \emph{epsilon expressivity} (Fig. \ref{Fig3}).

\subsubsection*{Curve generation and selection.}

The process begins by generating a representative set of nonlinear transfer curves accessible by a device/network.

For each network configuration (architecture $[1,1]$ or $[1,1,1]$, and SYNEs per edge $d_n \in \{1,2,3,5,8\}$), we generate a population of candidate $I$--$V$ transfer curves by randomising the SYNE control parameters and evaluating the resulting input-output response using the trained digital twin. Each candidate curve was obtained by sweeping the input voltage $V_\mathrm{In}$ from -2\,V to 2\,V and sampling 600 points throughout this range, recording the output current $I_\mathrm{Out}$. Tuning voltage parameters were sampled independently for every curve, every layer, and every SYNE using a sparse, even grid, with gains varied between -0.5 to 3, matching our experiments. To suppress responses dominated by the noise floor which would count highly towards expressivity but are not computationally useful, we reject any candidate curve whose raw output range $\max(y)-\min(y)$ is below a fixed threshold
$\Delta y_{\min} = 0.02 \times 10^{-8} = 2 \times 10^{-10}$, computed before any mean subtraction. Sampling continues until $K_{\mathrm{keep}}=2000$ accepted curves are obtained.

\subsubsection*{Epsilon packing in PCA space}

To make packing computationally efficient and consistent across configurations, we embed the curve set into a fixed $32$-dimensional PCA space. Concretely, we form the data matrix $Y\in\mathbb{R}^{K\times T}$ from the kept curves, center it across curves, compute a rank-$32$ PCA, and take the PCA scores $Z\in\mathbb{R}^{K\times 32}$.
Pairwise distances are Euclidean in PCA-score space,
\begin{equation}
D_{ij}=\lVert Z_i-Z_j\rVert_2,\qquad i<j,
\end{equation}
with PCA-space diameter $\mathrm{diam}=\max_{i<j}D_{ij}$.

\paragraph{Epsilon sweep.}
We sweep $\varepsilon$ over a combined grid formed from (i) log-spaced values from a small-distance lower bound (the $0.0005$ quantile of positive $\{D_{ij}\}$) up to $d_{\max}=1.95\times \mathrm{diam}$, and (ii) a quantile grid up to the $0.999$ quantile of $\{D_{ij}\}$. We report both absolute $\varepsilon$ and the normalised axis $\varepsilon/\mathrm{diam}$.

\paragraph{Greedy packing number with random restarts.}
For each $\varepsilon$, we estimate the packing number as the size of an $\varepsilon$-separated subset (all selected pairs satisfy $D_{ij}>\varepsilon$) constructed by a greedy procedure. Given an ordering $\pi$ of $\{1,\dots,K\}$, we scan points in that order; when a point $i$ is selected, all points within distance $\le \varepsilon$ of $i$ are removed from further consideration. Because greedy packing is order-dependent, we repeat this for $N_\mathrm{start}=8$ random permutations and record the packing counts across restarts, taking the best packing counts across trials as our reported values.

\subsubsection*{`Soft diameter' expressivity metric: Measuring the diameter of the SYNE nonlinear curve sets embedded as a PCA space point cloud}
To quantify the breadth of the resulting curve set while reducing sensitivity to a single extreme outlier pair, we computed a `soft diameter' in curve space. Let $\{y_k(V_{\mathrm{In}})\}_{k=1}^{K}$ denote the kept curves for a given configuration, sampled on the common $V_{\mathrm{in}}$ grid. We first remove the per-curve DC offset to focus on shape,
\begin{equation}
\tilde{y}_k = y_k - \langle y_k \rangle,
\end{equation}
and treat each mean-subtracted curve $\tilde{y}_k$ as a vector in $\mathbb{R}^{T}$ (with $T=N_{\mathrm{VIN}}$). We then form all pairwise Euclidean distances $d_{ij}=\lVert \tilde{y}_i-\tilde{y}_j\rVert_2$ and aggregate them using a log-sum-exp `soft max':
\begin{equation}
\mathrm{softdiam}_{\tau} \;=\; \tau \,\log\!\left(\sum_{i<j}\exp\!\left(\frac{d_{ij}}{\tau}\right)\right),
\qquad \tau = 0.25.
\end{equation}
This definition approaches the hard diameter $\max_{i<j} d_{ij}$ as $\tau \rightarrow 0$, but for finite $\tau$ it assigns non-negligible weight to multiple large (near-maximum) distances rather than only the single farthest pair. In practice, this makes $\mathrm{softdiam}_{\tau}$ a more useful measure which is less prone to distortion from a small number of very diverse/far-spaced points in an otherwise less diverse nonlinear curve set: it still increases when the most extreme separations grow, but it increases more robustly when many curve pairs are widely separated, rather than being dominated by one rare outlier. Figure \ref{SI-epsilon-vs-soft-diam} shows that for the devices and networks considered here, the soft-diameter follows a very similar trend to the epsilon packing threshold, and is quicker to evaluate - though this relation between the two metrics may not hold for all systems. We also compare the 50\% epsilon-packing threshold to the 95\% threshold, and show that similar trends hold for both.

Future comparative assessments across a broad variety of methods for assessing device/system level expressivity, including spectral approaches\cite{caravelli2025spectral}, and comparing multiple emerging physical devices, will be a valuable contribution to the field.

\subsection*{Pruning SYNEs from Synapses and Network Compression}

In all cases explored in this work, we begin by initialising networks with an equal number of SYNE devices per synapse. However, this can be inefficient. A network may not perform well at a task until synapses are initialised with a given number of SYNE devices, but this may be dominated by a small number of synapses (or even just one) where a complex function with high harmonic content is required, even though other synapses may need simpler functions where fewer SYNE devices would be sufficient, e.g. the comparison between function representation vs. number of SYNE devices shown in Figure \ref{Fig2}a) - perhaps only one synapse has a function complexity like the $\cos(2\pi x)$ which requires more than 2 SYNE devices, while the rest of the network has functions closer to the $ReLU(x)$ or $sigmoid(5x)$ for which 2 SYNEs is sufficient.

To address this, and reduce network size while minimising degradation of performance, we implement iterative pruning of SYNE devices from synapses. The methodology is simple - beginning from the trained network state with an equal, maximal number of SYNE devices on each synapse, we iteratively remove each SYNE device in turn, and evaluate test performance. So if the total maximal number of SYNE devices is $n$, we now run $n$ tests where in each test, the total number of SYNE devices in the network is $n-1$. As test evaluations require no training and physical KAN networks are compact, this is fast and computationally light. For each test, we evaluate the change in performance (e.g. MSE for regression tasks, accuracy for classification, etc) relative to the original un-pruned network state, and then rank each SYNE device by how much its removal degrades (or boosts!) network performance. We find that for a small number of SYNE devices per network (typically 1-3\% of total devices), removing the device actually improves network performance, either as the amplitude/range of the SYNE signal is relatively low, so signal-to-noise is poor, or because model-reality gap between the digital twin and experimental device means a given SYNE is not doing a good job of implementing the desired nonlinear function shape.

We then prune the network, successively removing SYNE devices based on their ranking and retraining the linear scaling and bias terms after each SYNE device is pruned, and evaluate performance vs. network size after each stage of device pruning. We can look at which synapses devices are being pruned from and which synapses are entirely pruned from the network, to evaluate relative importance of interactions between input features and relative complexity of nonlinear relationships within the network.

\subsection*{NASA Li-Ion Battery Charging and Aging Prediction}

This task was performed on a data set provided by NASA (https://data.nasa.gov/dataset/li-ion-battery-aging-datasets), and was collected from a custom built battery prognostics testbed at the NASA Ames Prognostics Center of Excellence (PCoE). Li-ion batteries were run through 3 different operational profiles (charge, discharge and Electrochemical Impedance Spectroscopy) at different temperatures. Discharges were carried out at different current load levels until the battery voltage fell to preset voltage thresholds. Some of these thresholds were lower than that recommended by the OEM (2.7 V) in order to induce deep discharge aging effects. Repeated charge and discharge cycles result in accelerated aging of the batteries. The experiments were stopped when the batteries reached the end-of-life (EOL) criteria of 30\% fade-in rated capacity (from 2 Ah to 1.4 Ah). Nine sensors/input values are output from the battery system: Six instantaneous measurements - battery temperature, voltage and current measured at the battery, voltage and current measured at the load, and time since the last charge; and three mean values taken as the mean between beginning the discharge cycle and the present measurement: battery temperature, voltage measured at the battery, and voltage measured at the load. 

The dataset contains 33 separate battery devices, and 2766 charge/discharge cycles. We sample 50 instantaneous measurements from each battery/discharge run to give training, validation and test sets of 96,810, 13,830 and 27,660 points respectively. A 10\% validation set is drawn from the training set. Training is run for 800 epochs using Adam optimiser, with a learning rate of 1e-4, and a cosine annealing scheduler.

\subsection*{MLP baselines}

To benchmark against conventional software networks, we trained and evaluated fully-connected multilayer perceptrons (MLPs) on each considered task. Each dataset separated into training and test splits: $X_\mathrm{Train}$, $X_\mathrm{Test}$, $y_\mathrm{Train}$, $y_\mathrm{Test}$. The test split was held out throughout training and used only for final evaluation.

\paragraph{Architecture sweep.}
For each task we trained a grid of ReLU MLPs with fixed hidden width $h \in \{50,75,100,150,200,250,300\}$ and hidden-layer depth $d \in \{1,2,3,4,5\}$ (maximum number of hidden layers evaluated was varied with task complexity), where $d$ denotes the number of hidden layers. All hidden layers had the same width.
ReLU nonlinearities are employed on hidden neurons. Linear layers are initialised with Xavier uniform weights and zero biases. The total parameter count $N$ was computed as the total number of trainable scalar weights and biases.

\paragraph{Training protocol and model selection.}
For each task, the provided training set was further split into a training subset and a validation subset by randomly selecting $10\%$ of training examples for validation (fixed split per task). Models were trained for 800 epochs using the Adam optimiser with fixed hyperparameters (selected in a separate hyperparameter search and then frozen for all tasks and model sizes): learning rate $\eta = 1.66\times 10^{-4}$, weight decay $3.05\times 10^{-5}$, and Adam betas $(\beta_1,\beta_2)=(0.916,0.972)$. The loss function was binary cross-entropy with logits (BCEWithLogitsLoss) for classification, and MSE loss for regression tasks and battery prediction. 

A cosine annealing learning-rate scheduler was applied using with $T_0=810$, $T_{\mathrm{mult}}=1$, and $\eta_{\min}=0.0858\,\eta$. Since $T_0$ exceeded the 800 training epochs, no restart occurred within a run and the schedule acted as monotonic cosine annealing toward $\eta_{\min}$. The model checkpoint used for testing was selected as the epoch with the lowest validation loss.

\paragraph{Trials, seeding, and hardware.}
To estimate optimisation variance, each (task, $h$, $d$) configuration was repeated for 15 independent trials with different random seeds affecting weight initialisation and minibatch order. Experiments were run in PyTorch on an Nvidia 4090 GPU, 

\label{Bibliography}
\bibliography{Bibliography.bib}

\newpage
\appendix

\setcounter{figure}{0}

\renewcommand{\thefigure}{S\arabic{figure}}

\captionsetup[figure]{labelformat=default,labelsep=period,name={Supplementary Figure}}

\newcommand{\figref}[1]{Fig.~S\ref{#1}}

\section*{Supplementary Information}

\subsection*{Tables comparing parameter and device counts between physical KANs and MLPs}

Tables comparing physical KAN (SYNE based) trainable parameter and device counts, and MLP trainable parameter and device counts are shown.

\begin{table}[t]
\centering
\small
\begin{tabular}{lrrrrrr}
\toprule
Architecture & SYNEs/synapse & Synapses & Nonlinear SYNE devices & Linear control devices & Total devices & Trainable params \\
\midrule
\texttt{[2,1,1]} & 6  & 3  & 18  & 54  & 72  & 92  \\
\texttt{[2,1,1]} & 8  & 3  & 24  & 72  & 96  & 122 \\
\texttt{[2,1,1]} & 12 & 3  & 36  & 108 & 144 & 182 \\
\texttt{[2,2,1]} & 6  & 6  & 36  & 108 & 144 & 183 \\
\texttt{[2,2,1]} & 8  & 6  & 48  & 144 & 192 & 243 \\
\texttt{[2,2,1]} & 12 & 6  & 72  & 216 & 288 & 363 \\
\texttt{[2,4,1]} & 6  & 12 & 72  & 216 & 288 & 365 \\
\texttt{[2,4,1]} & 8  & 12 & 96  & 288 & 384 & 485 \\
\texttt{[2,4,1]} & 12 & 12 & 144 & 432 & 576 & 725 \\
\texttt{[2,6,1]} & 6  & 18 & 108 & 324 & 432 & 547 \\
\texttt{[2,6,1]} & 8  & 18 & 144 & 432 & 576 & 727 \\
\texttt{[2,6,1]} & 12 & 18 & 216 & 648 & 864 & 1087 \\
\bottomrule
\end{tabular}
\caption{SYNE-KAN resource counts assuming per-SYNE tuning and gain voltages are realized by three additional linear memristive VCM devices per SYNE to provide non-volatile retention of trained parameters. Trainable parameters remain $5$ per SYNE plus one per hidden/output neuron (as previously defined).}
\end{table}

\begin{table}[t]
\centering
\small
\begin{tabular}{rrrr}
\toprule
$L$ & $h$ & Linear params/devices (weights) & Total params/devices \\
\midrule
1 & 50  & 150    & 201    \\
1 & 100 & 300    & 401    \\
1 & 200 & 600    & 801    \\
1 & 300 & 900    & 1201   \\
2 & 50  & 2650   & 2751   \\
2 & 100 & 10300  & 10501  \\
2 & 200 & 40600  & 41001  \\
2 & 300 & 90900  & 91501  \\
3 & 50  & 5150   & 5301   \\
3 & 100 & 20300  & 20601  \\
3 & 200 & 80600  & 81201  \\
3 & 300 & 180900 & 181801 \\
4 & 50  & 7650   & 7851   \\
4 & 100 & 30300  & 30701  \\
4 & 200 & 120600 & 121401 \\
4 & 300 & 270900 & 272101 \\
5 & 50  & 10150  & 10401  \\
5 & 100 & 40300  & 40801  \\
5 & 200 & 160600 & 161601 \\
5 & 300 & 360900 & 362401 \\
\bottomrule
\end{tabular}
\caption{MLP resource counts for 2 inputs, $L$ hidden layers of width $h$, and 1 output. Linear devices are weights; total params/devices equals weights plus biases (one bias per hidden/output neuron).}
\end{table}

\subsection*{Supplementary Note: Discussion of Potential Physical Mechanisms for the Electrical Transport Dynamics}
The P concentration used here ($9\times10^{17}$~cm$^{-3}$) is below the bulk Mott transition ($\gtrsim 3.7\times10^{18}$~cm$^{-3}$)~\cite{Rosenbaum_1980}. Prior work on comparably doped SOI films of similar thickness reports bulk-like dopant activation and near-complete ionization at room temperature \cite{pulici_electrical_2023}. In this regime, overlap of donor wavefunctions can lead to an impurity band within the Si band gap. Literature estimates place this band tens of meV below the conduction band edge with a meV-scale width for similar doping levels\cite{altermatt_simulation_2006}.

Nonlinear $I$--$V$ responses have previously been reported in nanometre-scale silicon devices with similar contact structure at low temperature and low readout bandwidth, and have been discussed in terms of variable-range hopping through dopant networks modulated by control electrodes\cite{chen_classification_2020,zolfagharinejad2025analogue,ruiz2021dopant,ruiz2020deep}. While related physics may contribute, our micrometre-scale devices exhibit strong nonlinearity at room temperature and MHz-rate operation, suggesting that additional mechanisms associated with thin SOI electrostatics and contact- and interface defects- induced depletion may play an important role as well as suggested in our recent work\cite{pulici_arxiv_2026}.  Consistent with this, analysis of input, output, and control currents indicates that SYNE devices can be viewed as a network of effective resistances between terminals that are strongly modulated by locally-applied bias voltages. Building and verifying a complete physical model describing the observed transport dynamics is outside the scope of this work and is fertile ground for future studies. However, irrespective of microscopic origin, the key property exploited here is that the applied tuning voltages provide deterministic and reproducible control over the nonlinear $I$--$V$ shape, including NDR.

\subsection*{List of functions used in function regression tasks}

\subsubsection*{Thirteen $f(x)$ functions used in Fig. 2d)}

$$
\cos(\pi x)
$$

$$
\sin(\pi x)
$$

$$
\lvert x\rvert \cos(2\pi x)
$$

$$
\sin^{2}(\pi x)
$$

$$
e^{-9x^{2}}
$$

$$
\cos^{2}(\pi x)
$$

$$
\frac{1}{1+e^{-5x}}
$$

$$
\frac{5x}{1+e^{-5x}}
$$

$$
\max(0,x)~or~ReLU(x)
$$

$$
x^{4}
$$

$$
x^{3} - 3x
$$

$$
\begin{cases}
1, & x = 0,\\[4pt]
\dfrac{\sin(10x)}{10x}, & x \neq 0
\end{cases}
~~or~\mathrm{sinc}(10x)
$$

$$
\arctan(5x)
$$

Numpy:

\texttt{lambda x: np.cos(np.pi * x)}\\
\texttt{lambda x: np.sin(np.pi * x)}\\
\texttt{lambda x: np.abs(x) * np.cos(2 * np.pi * x)}\\
\texttt{lambda x: np.sin(np.pi * x)**2}\\
\texttt{lambda x: np.exp(-9 * x**2)}\\
\texttt{lambda x: np.cos(np.pi * x)**2}\\
\texttt{lambda x: 1 / (1 + np.exp(-5 * x))}\\
\texttt{lambda x: (5 * x) / (1 + np.exp(-5 * x))}\\
\texttt{lambda x: np.maximum(0, x)}\\
\texttt{lambda x: x**4}\\
\texttt{lambda x: x**3 - 3*x}\\
\texttt{lambda x: np.where(10*x == 0, 1.0, np.sin(10*x) / (10*x))}\\
\texttt{lambda x: np.arctan(5*x)}\\

\subsubsection*{Six $f(g(x))$ functions used in Fig. 2c,e)}

Below are the six $f(g(x))$ functions and corresponding numpy code:
$$
\cos^{2}\!\left(e^{x}\right)
$$

$$
\cos(\pi x)\, e^{-\cos^{2}(\pi x)}
$$

$$
\lvert x^{3}\rvert \, \cos\!\left(2\pi x^{3}\right)
$$

$$
\arctan\!\Bigl(\pi\bigl(x^{3} - e^{-\lvert x\rvert}\cos(5x^{2})\bigr)\Bigr)
$$

$$
\operatorname{sgn}\!\bigl(\cos^{2}(\pi x)-\cos^{2}(x)\bigr)\,
\Bigl\lvert \cos^{2}(\pi x)-\cos^{2}(x)\Bigr\rvert^{1/3}
$$

$$
\max\!\bigl(0,\cos(\pi x)\bigr) ~ or ~\!\bigl(ReLU(\cos(\pi x)\bigr)
$$

Numpy functions:

\texttt{lambda x: (np.cos(np.exp(x))**2)}\\
\texttt{lambda x: (np.cos(np.pi * x)) * np.exp(-((np.cos(np.pi * x))**2))}\\
\texttt{lambda x: np.abs(x**3) * np.cos(2 * np.pi * (x**3))}\\
\texttt{lambda x: np.arctan(np.pi * ((x**3) - (np.exp(-np.abs(x)) * np.cos(5 * x**2))))}\\
\texttt{lambda x: np.sign((np.cos(np.pi*x)**2) - (np.cos(x)**2)) * np.abs((np.cos(np.pi*x)**2) - (np.cos(x)**2))**(1/3)}\\
\texttt{lambda x: np.maximum(0.0, np.cos(np.pi * x))}\\

\subsubsection*{Functions used for comparison against epsilon expressivity, Fig. 3c)}

$$
\sqrt[3]{\cos^{2}(\pi x)-\cos^{2}(x)}
$$

$$
\lvert x^{3}\rvert \,\cos\!\left(2\pi x^{3}\right)
$$

$$
\begin{cases}
1, & x = 0,\\[4pt]
\dfrac{\sin(10x)}{10x}, & x \neq 0
\end{cases}
$$

$$
\cos^{2}(\pi x)
$$

$$
\cos(2\pi x)
$$

$$
\sin^{2}(\pi x)
$$

$$
\lvert x\rvert \cos(2\pi x)
$$

$$
\cos(\pi x)\, e^{-\cos^{2}(\pi x)}
$$

$$
\arctan\!\Bigl(\pi\bigl(x^{3} - e^{-\lvert x\rvert}\cos(5x^{2})\bigr)\Bigr)
$$

$$
\max\!\bigl(0,\cos(\pi x)\bigr)
$$

\texttt{lambda x: np.sign((np.cos(np.pi*x)**2) - (np.cos(x)**2)) * np.abs((np.cos(np.pi*x)**2) - (np.cos(x)**2))**(1/3)}\\
\texttt{lambda x: np.abs(x**3) * np.cos(2 * np.pi * (x**3))}\\
\texttt{lambda x: np.where(10*x == 0, 1.0, np.sin(10*x) / (10*x))}\\
\texttt{lambda x: np.cos(np.pi * x)**2}\\
\texttt{lambda x: np.cos(2 * np.pi * x)}\\
\texttt{lambda x: np.sin(np.pi * x)**2}\\
\texttt{lambda x: np.abs(x) * np.cos(2 * np.pi * x)}\\
\texttt{lambda x: (np.cos(np.pi * x)) * np.exp(-((np.cos(np.pi * x))**2))}\\
\texttt{lambda x: np.arctan(np.pi * ((x**3) - (np.exp(-np.abs(x)) * np.cos(5 * x**2))))}\\
\texttt{lambda x: np.maximum(0.0, np.cos(np.pi * x))}\\






\begin{figure}[t!]
\centering
\includegraphics[width=1.0\textwidth]{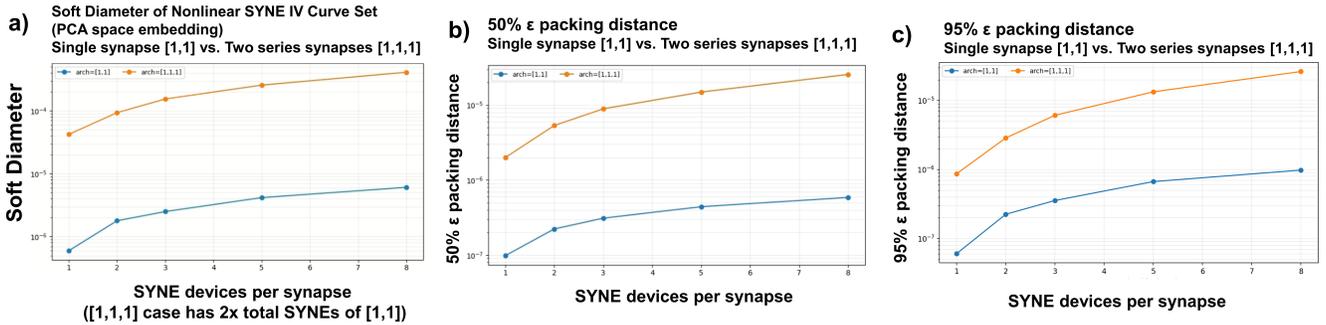}
\caption{\textbf{Comparing function representation performance for single vs. series simulated physical KAN synapses.} Mean MSE across 10 $f(x)$ and $f(g(x))$ functions is compared against parameter count for single [1,1] and two interconnected series [1,1,1] simulated physical KAN synapses (simulated via digital twin). Two interconnected synapses match or outperform equivalently-parameterised single synapses, and exhibit a somewhat steeper MSE vs. parameter count gradient.
}
\label{SI-epsilon-vs-soft-diam} 
\end{figure}

\begin{figure}[t!]
\centering
\includegraphics[width=0.94\textwidth]{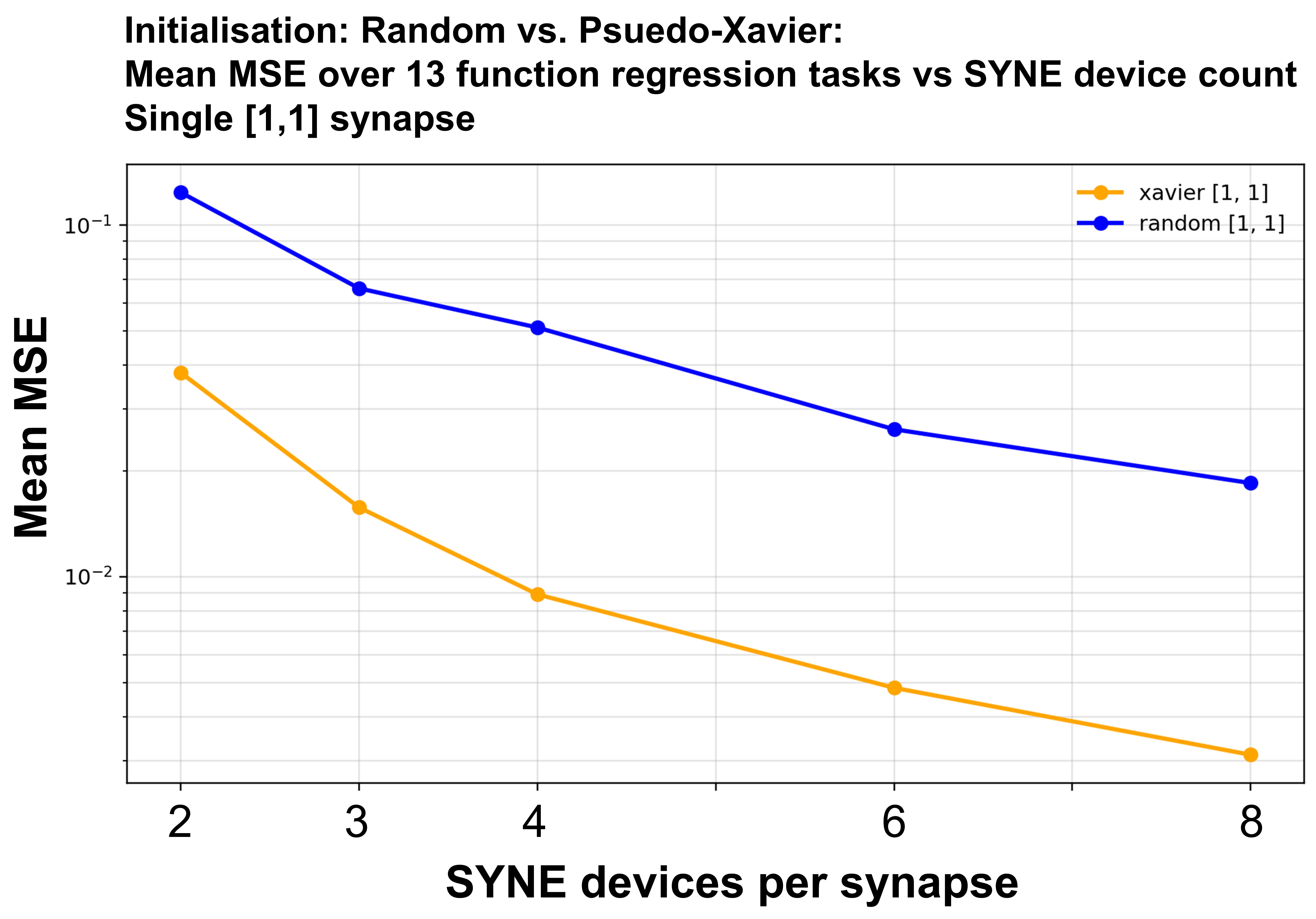}
\caption{\textbf{Random vs Xavier-style initialisation} Comparison of mean performance over 10 function regression tasks, between purely random parameter initialisation and Xavier-style initialisation. Xavier-style initialisation is found to give consistently better performance.}
\label{SI-Random-vs-Xav} 
\end{figure}

\begin{figure}[t!]
\centering
\includegraphics[width=1.0\textwidth]{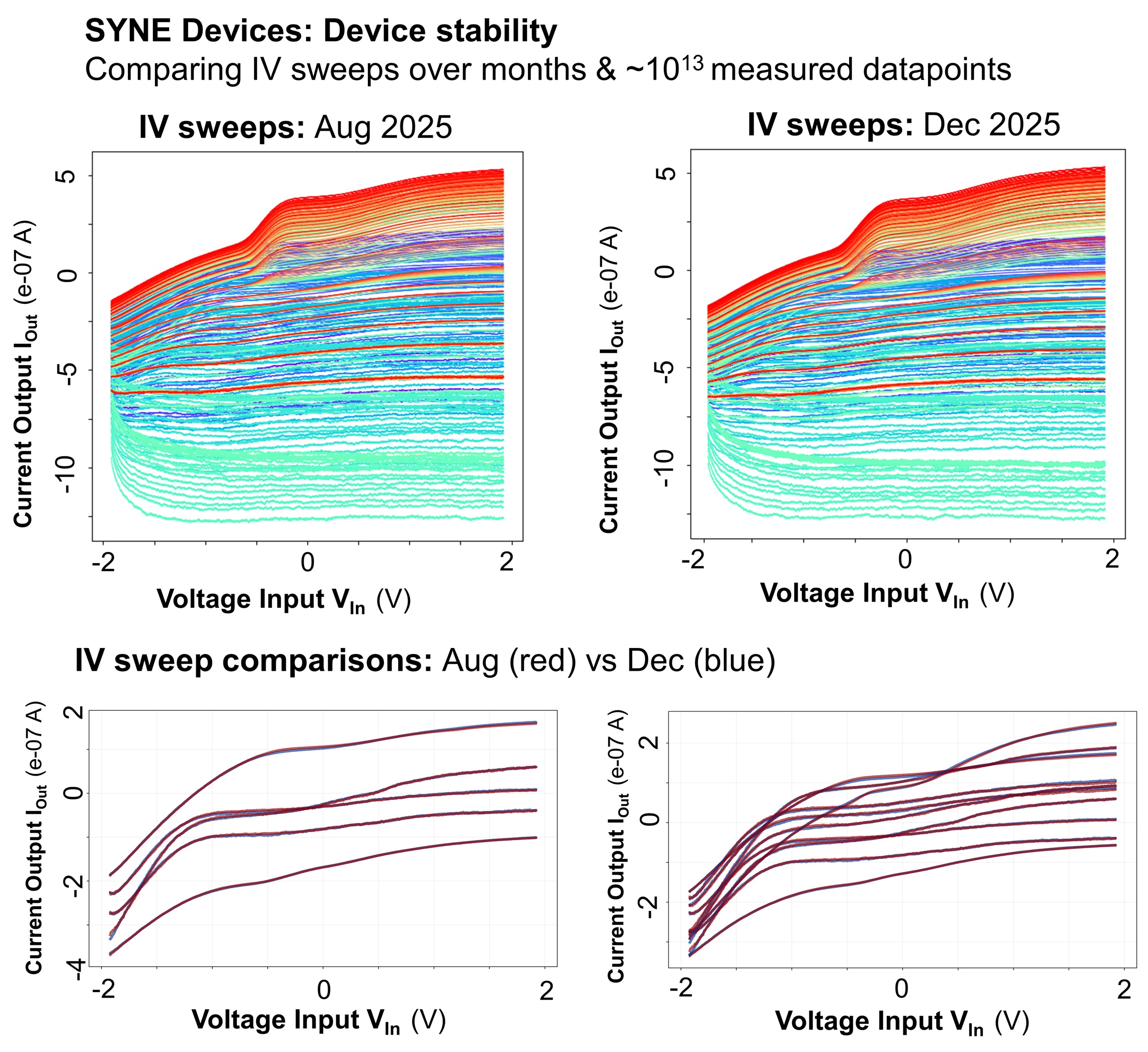}
\caption{\textbf{Device stability} Comparison of SYNE $I$--$V$ traces measured months apart, with approximately $10^{13}$ data points measured in between the $I$--$V$ trace sets shown here. No significant device degradation or variation in response is observed.
}
\label{SI-device-stability} 
\end{figure}

\begin{figure}[t!]
\centering
\includegraphics[width=0.94\textwidth]{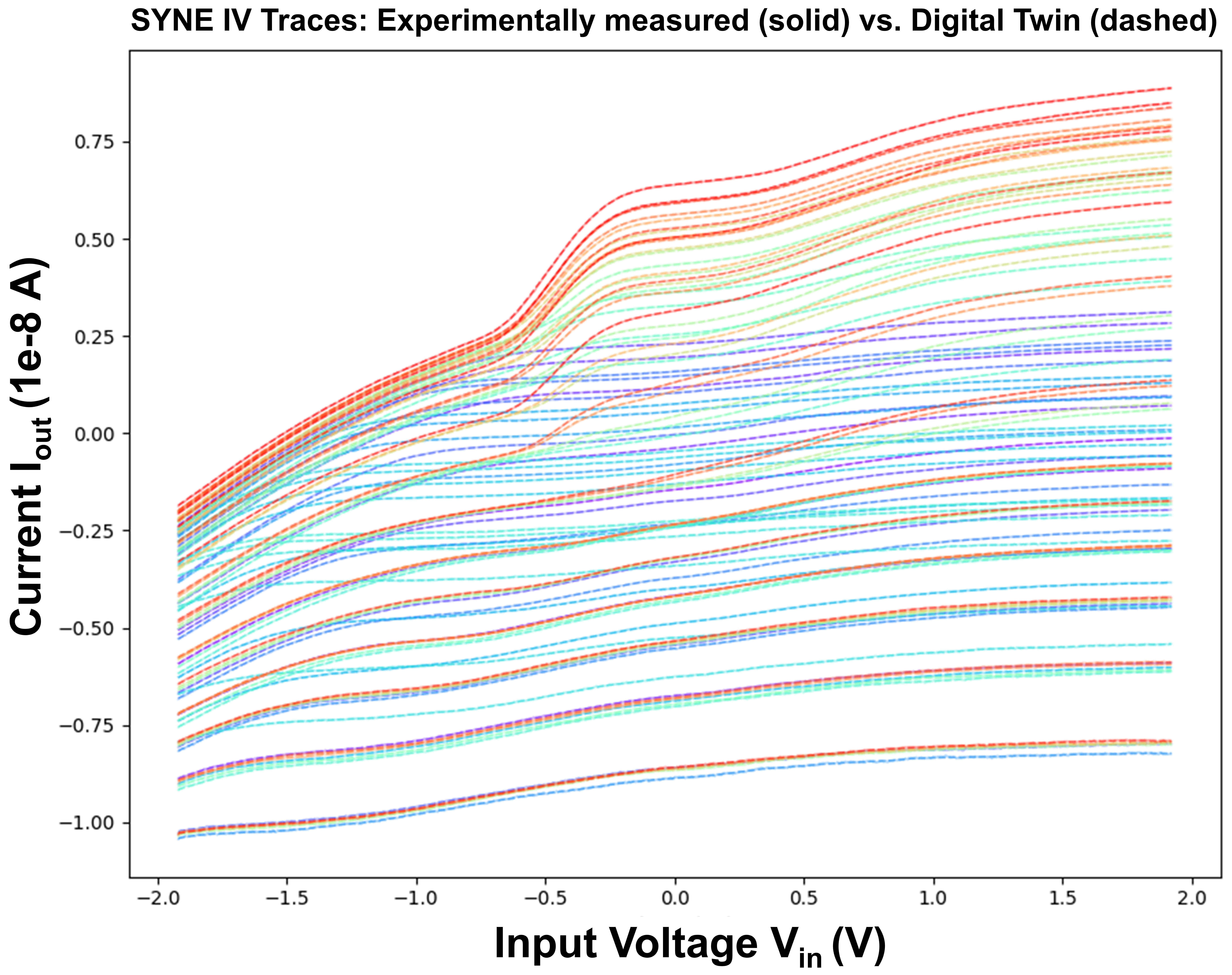}
\caption{\textbf{Confirmation of Digital Twin Performance} Digital twin output compared against experimentally measured $I$--$V$ traces, showing good reproduction of experimental behaviour by the digital twin.}
\label{SI-Digi-Twin-Test} 
\end{figure}

\begin{figure}[t!]
\centering
\includegraphics[width=0.94\textwidth]{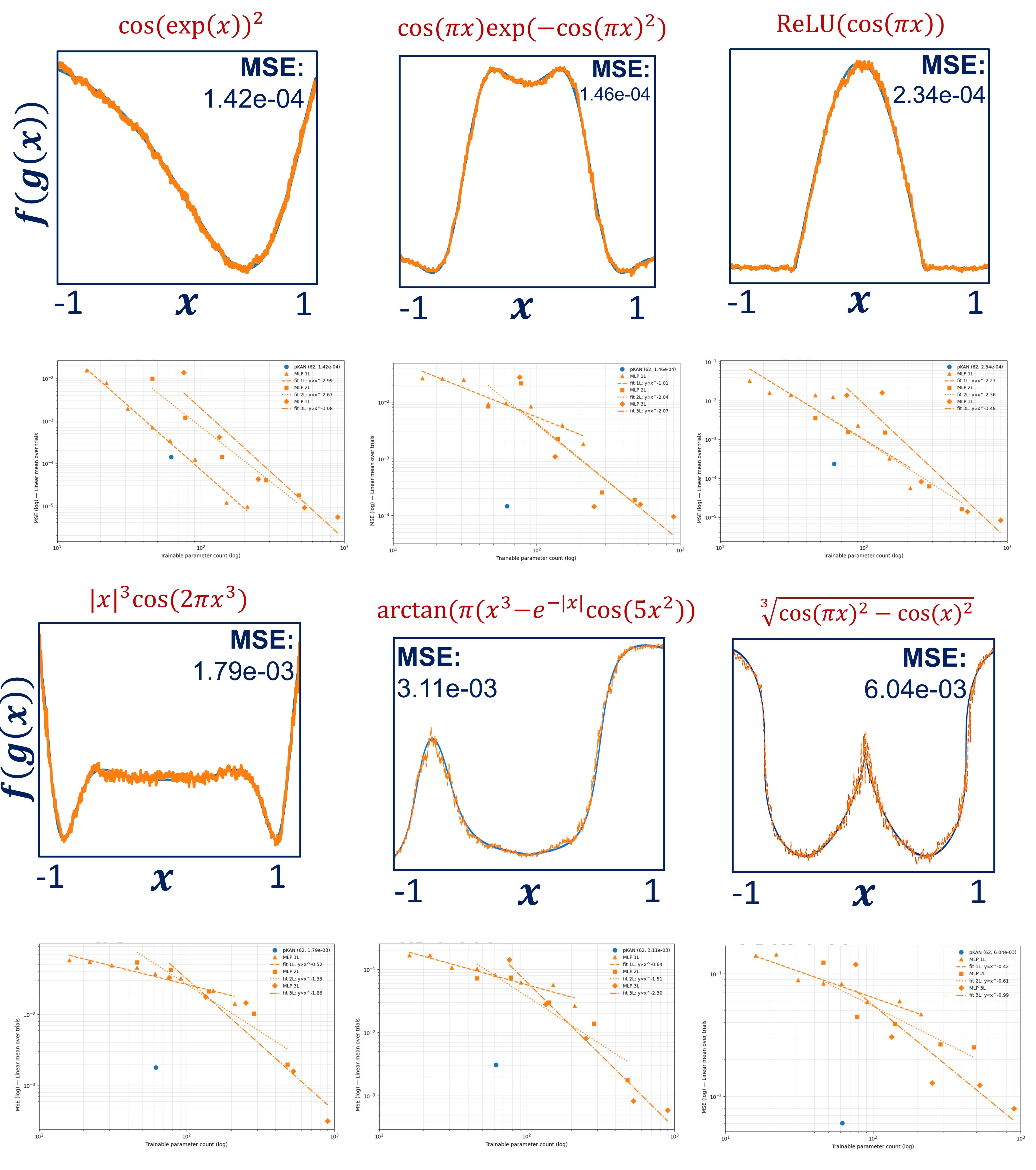}
\caption{\textbf{Nested f(g(x)) function regression via two series-connected physical KAN synapses and MLP baselines} Separate MSE vs. parameter count plots for each of the six f(g(x)) functions in Figs. 2c) and 2e).}
\label{SI-f-g-x-function-regression} 
\end{figure}

\vspace{28pt}


\end{document}